%% file: main.tex
\documentclass[modern]{aastex63}
\usepackage{threeparttable}
\usepackage{xspace}

\newcommand{\ktwo}{\textit{K2}}
\newcommand{\kms}{km~s$^{-1}$\xspace}

\newcommand{\masyr}{mas~yr$^{-1}$}

\newcommand{\teff}{$T_\mathrm{eff}$}
\newcommand{\msun}{$M_\odot$}
\newcommand{\rsun}{$R_\odot$}

\newcommand{\rstar}{$R_*$}

\newcommand{\vrad}{$v_{R}$}
\newcommand{\pmra}{$\mu_{\alpha}$}
\newcommand{\pmdec}{$\mu_{\delta}$}

\newcommand{\mjup}{$M_\mathrm{Jup}$}

\newcommand{\gaia}{\textit{Gaia}}
\newcommand{\kepler}{\textit{Kepler}}

\mathchardef\mhyphen="2D

\shorttitle{Asymmetric Eclipse of V928 Tau}

\begin{document}

\correspondingauthor{Dirk M. van Dam}
\email{dmvandam@strw.leidenuniv.nl}

\author[0000-0002-1033-3461]{Dirk M. van Dam}
\affil{Leiden Observatory, Leiden University, P.O. Box 9513, 2300 RA Leiden, The Netherlands}

\author[0000-0002-7064-8270]{Matthew A. Kenworthy}
\affil{Leiden Observatory, Leiden University, P.O. Box 9513, 2300 RA Leiden, The Netherlands}

\author[0000-0001-6534-6246]{Trevor J.\ David}
\affil{Center for Computational Astrophysics, Flatiron Institute, New York, NY 10010, USA}
\affil{Jet Propulsion Laboratory, California Institute of Technology, 4800 Oak Grove Drive, Pasadena, CA 91109, USA}

\author[0000-0003-2008-1488]{Eric E.\ Mamajek}
\affil{Jet Propulsion Laboratory, California Institute of Technology, 4800 Oak Grove Drive, Pasadena, CA 91109, USA}
\affil{Department of Physics \& Astronomy, University of Rochester, Rochester, NY 14627, USA}

\author{Lynne A.\ Hillenbrand}
\affil{Department of Astronomy, California Institute of Technology, Pasadena, CA 91125, USA}

\author[0000-0002-3656-6706]{Ann Marie Cody}
\affil{Bay Area Environmental Research Institute, 625 2nd Street, Ste. 209, Petaluma, CA 94952, USA}

\author[0000-0001-8638-0320]{Andrew W.\ Howard}
\affil{Department of Astronomy, California Institute of Technology, Pasadena, CA 91125, USA}

\author[0000-0002-0531-1073]{Howard Isaacson}
\affiliation{501 Campbell Hall, University of California at Berkeley, Berkeley, CA 94720, USA}
\affiliation{Centre for Astrophysics, University of Southern Queensland, Toowoomba, QLD, Australia}

\author[0000-0002-5741-3047]{David R.\ Ciardi}
\affil{California Institute of Technology/IPAC-NASA Exoplanet Science Institute, Pasadena, CA 91125, USA}

\author[0000-0001-6381-515X]{Luisa M.\ Rebull}
\affil{Infrared Science Archive (IRSA), IPAC, 1200 E. California Blvd., California Institute of Technology, Pasadena, CA 91125, USA}

\author[0000-0003-3595-7382]{John R.\ Stauffer}
\affil{Spitzer Science Center (SSC), IPAC, 1200 E. California Blvd., California Institute of Technology, Pasadena, CA 9112, USA}

\author[0000-0002-5025-6827]{Rahul Patel}
\affil{Infrared Processing and Analysis Center, California Institute of Technology, Pasadena, CA 91125, USA}

\author[0000-0002-8863-7828]{Andrew Collier Cameron + WASP Collaborators}
\affil{Centre for Exoplanet Science, SUPA, School of Physics and Astronomy, University of St Andrews, St Andrews KY16 9SS, UK}

\author[0000-0001-8812-0565]{Joseph E. Rodriguez}
\affil{Center for Astrophysics \textbar\ Harvard \& Smithsonian, 60 Garden St, Cambridge, MA 02138, USA}

\author[0000-0002-6495-0676]{Grzegorz Pojma\'nski}
\affil{Astronomical Observatory, University of Warsaw, 
Al. Ujazdowskie 4, 00-478 Warszawa,Poland}

\author[0000-0002-9329-2190]{Erica J.\ Gonzales}
\affil{Department of Astronomy \& Astrophysics, University of California, Santa Cruz, CA 95064, USA}
\altaffiliation{National Science Foundation Graduate Research Fellow}

\author[0000-0001-5347-7062]{Joshua E.\ Schlieder}
\affil{NASA Goddard Space Flight Center, 8800 Greenbelt Rd, Greenbelt, MD 20771, USA}


\author{Franz-Josef Hambsch}
\affil{American Association of Variable Star Observers (AAVSO), 49 Bay State Road, Cambridge, MA 02138, USA}
\affil{Vereniging Voor Sterrenkunde (VVS), Oostmeers 122 C, 8000 Brugge, Belgium}

\author{Sjoerd Dufoer}
\affil{Vereniging Voor Sterrenkunde (VVS), Oostmeers 122 C, 8000 Brugge, Belgium}

\author{Tonny Vanmunster}
\affil{American Association of Variable Star Observers (AAVSO), 49 Bay State Road, Cambridge, MA 02138, USA}
\affil{CBA Belgium Observatory, Walhostraat 1A, B-3401 Landen, Belgium}
\affil{CBA Extremadura Observatory, 06340 Fregenal de la Sierra, Badajoz, Spain}

\author{Franky Dubois}
\affil{American Association of Variable Star Observers (AAVSO), 49 Bay State Road, Cambridge, MA 02138, USA}
\affil{Vereniging Voor Sterrenkunde (VVS), Oostmeers 122 C, 8000 Brugge, Belgium}
\affil{Astrolab IRIS, Verbrandemolenstraat, Ypres, Belgium}

\author[0000-0003-0231-2676]{Siegfried Vanaverbeke}
\affil{American Association of Variable Star Observers (AAVSO), 49 Bay State Road, Cambridge, MA 02138, USA}
\affil{Vereniging Voor Sterrenkunde (VVS), Oostmeers 122 C, 8000 Brugge, Belgium}
\affil{Astrolab IRIS, Verbrandemolenstraat, Ypres, Belgium}

\author{Ludwig Logie}
\affil{Astrolab IRIS, Verbrandemolenstraat, Ypres, Belgium}

\author{Steve Rau}
\affil{Astrolab IRIS, Verbrandemolenstraat, Ypres, Belgium}
\title{An Asymmetric Eclipse Seen Towards the Pre-Main Sequence\\ Binary System V928 Tau}

\begin{abstract}
\ktwo\ observations of the weak-lined T Tauri binary V928 Tau A+B show the detection of a single, asymmetric eclipse which may be due to a previously unknown substellar companion eclipsing one component of the binary with an orbital period $>$ 66 days.
Over an interval of about 9 hours, one component of the binary dims by around 60\%, returning to its normal brightness about 5 hours later.
From modeling of the eclipse shape we find evidence that the eclipsing companion may be surrounded by a disk or a vast ring system.
The modeled disk has a radius of $0.9923\,\pm\,0.0005\,R_*$, with an inclination of $56.78\,\pm\, 0.03^\circ$, a tilt of $41.22\,\pm\,0.05^\circ$, an impact parameter of $-0.2506\,\pm\,0.0002\,R_*$ and an opacity of 1.00.
The occulting disk must also move at a transverse velocity of $6.637\,\pm\,0.002\,R_*\,\mathrm{day}^{-1}$, which depending on whether it orbits V928 Tau A or B, corresponds to approximately 73.53 or 69.26 \kms.
A search in ground based archival data reveals additional dimming events, some of which suggest periodicity, but no unambiguous period associated with the eclipse observed by \ktwo.
We present a new epoch of astrometry which is used to further refine the orbit of the binary, presenting a new lower bound of 67 years, and constraints on the possible orbital periods of the eclipsing companion. 
The binary is also separated by 18\arcsec\ ($\sim$2250 au) from the lower mass CFHT-BD-Tau 7, which is likely associated with V928 Tau A+B. 
We also present new high dispersion optical spectroscopy that we use to characterize the unresolved stellar binary.
\end{abstract}

\keywords{Pre-main sequence stars (1290) --- Astrometric binary stars (79) ---  Eclipses (442) --- Planetary rings (1254) --- Substellar companion stars (1648)}

\section{Introduction}
\label{sec:intro}

With the advent of high precision photometric telescopes from the ground and space, astronomers have been able to continuously observe a large number of stars that exhibit intriguing behaviour in their apparent brightness.
This can come from intrinsic stellar variability, i.e. high amplitude optical variability of young stars \citep{Joy:1945}, rotational starspot modulation \citep{Rodono:etal:1986,Olah:etal:1997}, asteroseismology \citep{Handler:2013}; or from interactions with objects or dust orbiting the star.
`Dipper' stars are a class of stars where occultations due to dust in the inner boundaries of circumstellar disks produce transits with depths of up to 50\% \citep{Alencar:etal:2010,Cody:etal:2014,Cody_Hillenbrand:2018}.
\cite{Ansdell:etal:2019a} found that in some cases this requires misalignment of the inner protoplanetary disk compared to the circumstellar disk, and \cite{Ansdell:etal:2019b} found that shallower dips could be caused by exo-comets.
A particular system of interest, due to its evolved age, is RZ Psc studied by \cite{Kennedy:etal:2017}, which is a Sun-like star exhibiting transits of dust clumps, that could originate from an asteroid belt analogue of the Solar System.
Other intriguing transits include disintegrating planets, which have regular periods but varying transit depths due to the loss of planetary material \citep{Rappaport:etal:2012,Lieshout:2018,Ridden_Harper:2018}.

An additional source of deep asymmetric eclipses, which we will explore further in this study, is the transit of tilted and inclined circum-``planetary'' disks, which due to projection effects, create elliptical occulters.
These systems are interesting as they reveal the formation mechanism of planets, particularly if we observe young systems.
Circumstellar disks are a fundamental feature of stellar formation and the planets that form in these disks are influenced by the structure and composition of the protoplanetary disk, the interaction with the young host star and the different formation mechanisms of planets \citep[see reviews by][]{Armitage:2011,Kley:Nelson:2012}.
Direct imaging allows astronomers to study the general size, shape and composition of these circum-``planetary'' disks, but the transit method allows the spatial structure to be probed indirectly with a resolution significantly higher than through direct imaging.
Besides providing insight into planet formation, these systems also reveal the mechanisms of ring and moon formation \citep{Teachey:Kipping:Schmitt:2018}.
Other systems that have been explored include: EPIC 204376071 \citep{Rappaport:etal:2019}, 1SWASP J140745.93--394542.6J1407 \citep[J1407,][]{Kenworthy_2015} and PDS 110 \citep{Osborn2017,Osborn:etal:2019}.

The \kepler\ space telescope \citep{Kepler} was designed to determine the frequency of Earth-sized planets in and near the habitable zone of Sun-like stars, $\eta_{\mathrm{Earth}}$, which as a consequence produced a large number of high precision light curves. 
After the failure of the second of its four reaction wheels, the mission was reconfigured to the extended \ktwo\ mission \citep{Howell:etal:2014}, which observed fields along the ecliptic.
\ktwo\ has found several of these deep asymmetric eclipses, which have been compiled into a comprehensive list by \cite{LaCourse_2018}.
Here we present \ktwo\ observations of the pre-main sequence binary star V928 Tau which shows a deep and asymmetric eclipse, potentially due to a previously unknown companion orbiting one component of the binary.
The nature of the source of extinction is unknown, but consistent with a small dust disk. 

In Section~\ref{sec:star} we present and determine the properties of V928 Tau. 
Section ~\ref{sec:observations} describes all the observations of the system from photometry, spectroscopy, astrometry to high resolution imaging.
Section~\ref{sec:lc} describes all the analysis performed on the \ktwo\ light curve.
This includes the modelling of the stellar variation, the eclipse and a periodicity search.
We summarise and discuss our findings in Section~\ref{sec:discussion}.
The preliminary results for this system where presented in \citet{vanDam:etal:2019}.

\section{Stellar Characterization}
\label{sec:star}

\subsection{Literature}
\label{subsec:lit}
The current state of published knowledge about V928 Tau is summarized in Table~\ref{table:star} and in the succeeding subsections.
\input{table_star}
%

    \textit{Membership Provenance:} V928 Tau is a proposed member of the Taurus-Auriga star-forming complex ($d \sim$ 145 pc, $\tau \sim$ 0--5 Myr). 
    The star's membership was first proposed by \citet{Jones:Herbig:1979} on the basis of proper motions and it was given the designation JH 91. 
    Other aliases include L1529-23, EPIC 247795097 and HBC 398.
    %

    \textit{Environment:} V928 Tau is located in the TMC 2 region of the dark cloud complex B18 \citep[Kutner's cloud,][]{Leinert:etal:1993}, and belongs to the Tau IV subgroup \citep{Gomez:etal:1993, Luhman:etal:2009}. 
    The star is separated by 18.18\arcsec\ from another Tau-Aur member, CFHT-BD-Tau 7 (2MASS J04321786+2422149, EPIC 247794636), which resides in the same \ktwo\ postage stamp. 
    Statistical analysis of the spatial distribution of Taurus members suggests that stars this close (18.18\arcsec\, $\simeq$ 2250 au at a distance of 124 pc) are almost certainly physical multiples \citep{Gomez:etal:1993, Joncour:etal:2017, Joncour:etal:2018}. 
    Astrometric information on the environment of V928 Tau is summarised in Table~\ref{table:kin}.
    
    \input{table_kinematics.tex}

    \textit{Binarity:} V928 Tau was first discovered to be a binary through lunar occultation observations at 2.2 \micron\ and followed up with speckle imaging, which revealed the two stars to be closely separated on the sky ($\rho$ $\approx$ 0.2$\arcsec$) and nearly equal in brightness at $K$ band \citep{Leinert:etal:1993}. 
    \citet{Schaefer:etal:2014} analyzed the astrometric motion of the binary V928 Tau based on newly acquired Keck NIRC2 data and previously published measurements from the literature \citep{Leinert:etal:1993, Ghez:etal:1993, Ghez:etal:1995, Simon:etal:1996, White:Ghez:2001, Kraus:Hillenbrand:2012}.
    From the compilation of measurements, those authors found the projected motion of the binary could not be distinguished from linear motion.
    However, assuming the pair is bound, those authors found an orbital period greater than 58 years was required to fit the data.
    \citet{Kraus:Hillenbrand:2012} characterized the binary further, deriving a mass ratio of $q$ = 0.97, individual masses ($M_1=0.60$ \msun, $M_2=0.58$ \msun), and the projected separation (32 au).  
    The likely association with CFHT-BD-Tau 7 at $\simeq$ 2250 au has been proposed in \citet{Guieu:etal:2006} and explored further in \citet{Kraus:Hillenbrand:2009b}.
    The multiplicity is explored further in \citet{Joncour:etal:2018}, where V928 Tau and CFHT-BD-Tau 7 are in NEST 9.
    %
    
    \textit{Circumstellar Disk:} The star is a weak-lined T Tauri with modest H$\alpha$ emission \citep[EW(H$\alpha$) = -1.2 to -2.4 \AA,][this work]{Cohen:Kuhi:1979, Feigelson:Kriss:1983, Kenyon:etal:1998, Dent:etal:2013} and a class III spectral energy distribution \citep[$L_{\mathrm{FIR}} / L_{\mathrm{bol}}$$< 0.04$,][]{Kenyon:etal:1998}.
    The state of a putative disk has been studied numerous times over the years, beginning with \cite{Strom:etal:1989}. 
    Recently, \citet{Dent:etal:2013} estimated an upper limit to the mass of dust within the system of $< 4 \times 10^{-6}$ \msun.
    %

    \textit{Spectral Type:} 
    An initial spectral classification of M0.5 was determined for this star by \citet{Cohen:Kuhi:1979} and \citet{Feigelson:Kriss:1983} quoted K7/M0e.
    From a flux-calibrated low-resolution optical spectrum of V928 Tau, \citet{Herczeg:Hillenbrand:2014} determined a more precise combination of spectral type (M0.8), $V$-band extinction (1.95~mag), and veiling at 7510~\AA\ (0.00). 
    Those authors also used the \citet{Tognelli:etal:2011} evolutionary models to determine the stellar mass (0.5~\msun) and age (1.6~Myr), under the assumption of a single star. 
    \cite{Tottle:Mohanty:2015} fit model atmospheres to the spectral energy distribution of V928 Tau, finding \teff\ = 3525 K, $A_J = 0.94$ mag, and log $L/L_\odot = 1.04$.
    \citet{Kounkel:etal:2019} determined from an analysis of H-band spectra a somewhat warmer temperature of \teff\ = 4190 K, and log g = 4.31 cm s$^{-2}$ along with a veiling value at 1.6 $\mu$m of 0.11.
    From our Keck/HIRES spectra we derive a spectral type of K9.0 $\pm$ 0.9, which is between the M2 and the K6 that are implied by the two temperatures given above. 
    We ultimately adopt the M0.8 $\pm$ 0.5 found by \citet{Herczeg:Hillenbrand:2014} because, for M-type stars, spectral typing is considered more accurate at lower spectral resolution than higher. 
    From adaptive-optics resolved spectroscopy, V928 Tau A and B are found to have nearly identical near-infrared spectra (L. Prato, private communication). 
    Assuming the stars are in fact physically associated, the nearly identical spectra reinforce the notion that the two components have very similar bulk properties, such as mass and radius. 
    %

    \textit{Radial and Rotational Velocity:} 
    \citet{Hartmann:etal:1986} first measured the radial velocity (18.3~\kms) and $v\sin{i}$ (24.9~\kms) for V928 Tau.
    Next, from four epochs of seeing-limited, high-resolution spectroscopy, \citet{Nguyen:etal:2012} measured the radial velocity to be 15.38 $\pm$ 0.16 \kms\ (with a weighted standard deviation of 1.67 \kms\ and systematic noise of 2.02 \kms).
    Those authors also measured $v\sin{i}$ to be 31.6 $\pm$ 0.7 \kms.

    RV data including the previous as well as our three new measurements are summarised in Table~\ref{table:rvs}.
    Rotation data appear in Table~\ref{table:star}.
    Other rotation measurements, in addition to those above, include \citet{Hartmann:Stauffer:1989} that reported $v\sin{i}$ = 18.8 $\pm$ 3.3 \kms. 
    From our Keck/HIRES data we determine a $v\sin{i}$ = 29 $\pm$ 3 \kms\ for the first, 2017 epoch and 33.1 $\pm$ 1.2 \kms\ for the third, 2018 epoch.
    \citet{Kounkel:etal:2019} reported $v\sin{i}$ = 34.2 $\pm$ 0.4 \kms\ from APOGEE.
    We emphasize again that these measurements are for the combined (spatially unresolved) $A+B$ stellar system.
    We also note that the various $v\sin{i}$ measurements were acquired with different spectral resolutions: $\sim$~5~\kms \citep{Nguyen:etal:2012}, $\sim$~8~\kms (this work), and $\sim$12--13~\kms \citep{Hartmann:etal:1986, Hartmann:Stauffer:1989, Kounkel:etal:2019}. 
    For comparison, the maximum velocity separation between the components for an assumed orbital period of 60 years is 8~\kms. 
    \input{table_rvs}

\subsection{Reddening}
\label{subsec:reddening}
\citet{Herczeg:Hillenbrand:2014} measured the extinction towards V928 Tau from a flux-calibrated optical spectrum, finding $A_V = 1.95 \pm 0.2$~mag.
This value is consistent with a local, high-resolution extinction map \citep{Dobashi:etal:2005}.
Using the 2MASS extinction coefficients of \citet{Yuan:etal:2013} and assuming $R_V = 3.1$ \citep{Cardelli:etal:1989}, we calculated the extinction corrected near-infrared colors of the primary and secondary from the de-blended photometry: $(J-K)_\mathrm{0, pri} = 1.14 \pm 0.05$~mag and $(J-K)_\mathrm{0, sec} = 1.21 \pm 0.05$~mag. 

\subsection{Stellar Parameters}
\label{subsec:params}
From the veiling-corrected spectral type of \citet{Herczeg:Hillenbrand:2014} and its associated uncertainty, we determined $T_\mathrm{eff}$ via Monte Carlo error propagation and linear interpolation of Table 6 from \citet{Pecaut:Mamajek:2013}, appropriate for pre-main sequence stars. 
Using the same table and methods we determined the $J$ band bolometric correction, absolute $J$ magnitude, bolometric magnitude, luminosity, and radius for each star (assuming the two stars have equivalent effective temperatures). 
We then performed linear interpolation of the Dartmouth evolutionary models, both the standard \citet{Dotter:etal:2008} and magnetic \citet{Feiden:2016} versions, to determine masses and ages in the H-R diagram. 
Our derived stellar parameters are reported in Table~\ref{table:star}. 

\subsection{Stellar Radii}
\label{subsec:radii}
While it is not clear which component of the binary is being transited or eclipsed, or whether the multiple dips observed by \ktwo\ and ground-based surveys may in fact be due to separate companions around both stars, our analysis is simplified somewhat by the fact that the two stars in the binary are nearly identical. 
From the two obvious rotation periods detected from \ktwo\ photometry and the $v\sin{i}$ value published in \citet{Nguyen:etal:2012}, the minimum stellar radius can be calculated as $R_*\sin{i}$ = 1.41~\rsun\ or 1.56~\rsun, depending on which period is used (and neglecting differential rotation).

\section{Observations}
\label{sec:observations}
Here we summarise all the observations we collected on V928 Tau. Time series photometry from \ktwo\ and ground based surveys, spectroscopy from Keck-I HIRES, \gaia\ DR2 data and high-resolution imaging from Keck-II NIRC2.

\subsection{Time Series Photometry} \label{sec:photometry}

\subsubsection{K2}
\begin{figure}
    \centering
    \includegraphics[width=0.29\textwidth]{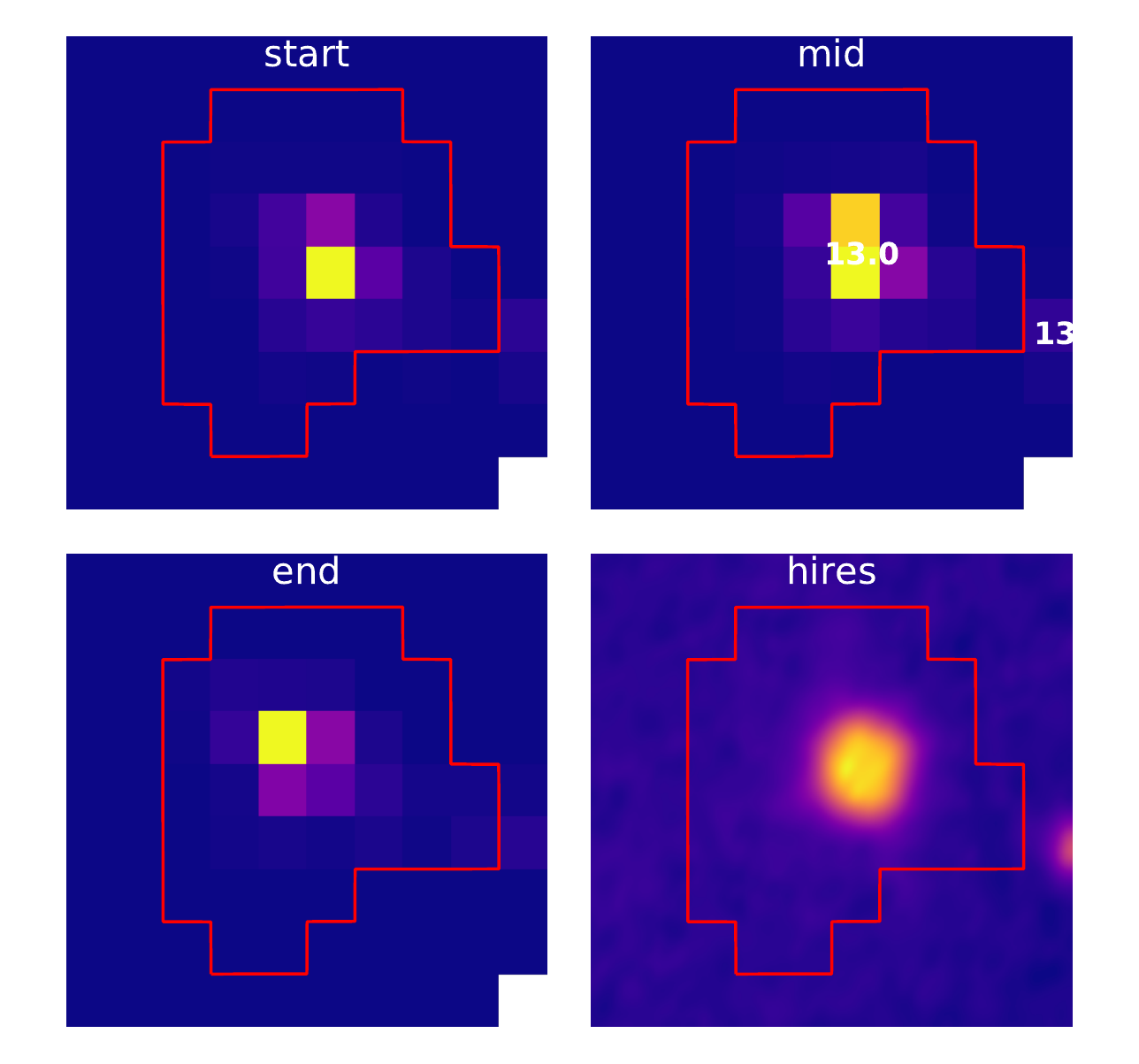}    
    \includegraphics[width=0.69\textwidth]{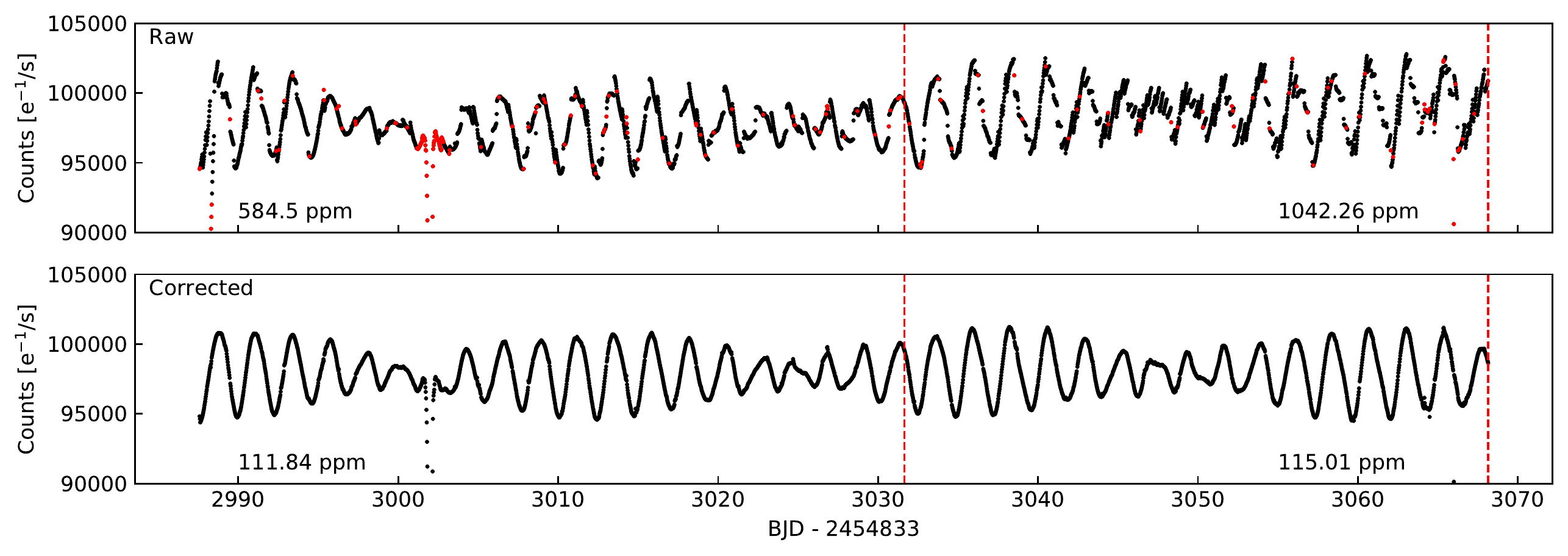}
    \caption{\textit{Left}: the aperture used to compute \ktwo\ light curve of V928 Tau with \texttt{EVEREST 2.0}. The high resolution image in the lower right panel is taken from the Palomar Observatory Sky Survey while the others images are from \ktwo. \textit{Right}: the raw (\textit{top}) and corrected (\textit{bottom}) \ktwo\ light curve of V928 Tau. Systematics were corrected using the Pixel Level Decorrelation (PLD) model of \texttt{EVEREST 2.0}. Red points were masked when computing the PLD model. Vertical red dashed lines indicate breakpoints. The Combined Differential Photometric Precision (CDPP) values on either side of the breakpoint are indicated in the lower portion of each panel. Note the eclipse occurring at BJD $\sim$ 2457835 with a depth of $\sim$ 30 \%.}
    \label{fig:everest-corrections}
\end{figure}

\begin{figure}
    \centering
    \includegraphics[width=\textwidth]{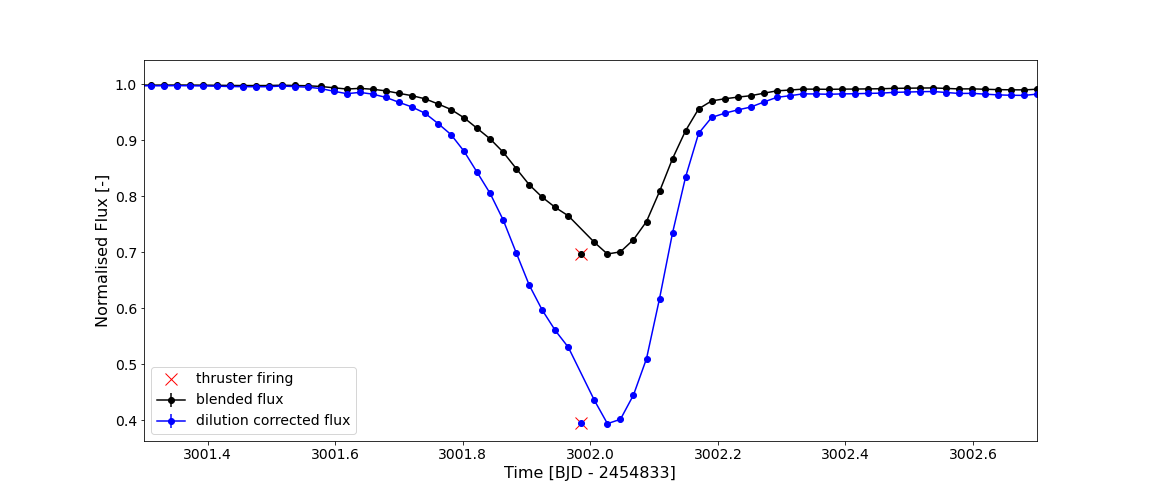}
    \caption{Light curve of V928 Tau after removing stellar variations centered on the eclipse. This figure shows the blended and dilution corrected eclipse based on the assumption of two identical stars. A single observation during the eclipse was excluded due to a thruster firing (marked with a red X).}
    \label{fig:eclipse}
\end{figure}

V928 Tau (EPIC 247795097) was observed by the \kepler\ space telescope between 2017-03-08 UT and 2017-05-27 UT during Campaign 13 of the \ktwo\ mission. 
The \ktwo\ light curve was extracted using the \texttt{EVEREST 2.0} pipeline \citep{Luger:etal:2016,Luger:etal:2018}, which uses a variant of Pixel Level Decorrelation (PLD) to correct for the systematics in the \citet{Vanderburg:Johnson:2014} light curves.
The light curve consist of 9344 observations, spanning $\sim$ 80 days, with a Combined Differential Photometric Precision (CDPP) of $\sim$ 113 ppm.
This light curve is characterized by quasi-periodic brightness modulations, a beating pattern and a deep asymmetric eclipse seen at BJD $\sim$ 2457835 (see Figure~\ref{fig:everest-corrections}).
Using the \texttt{lightkurve} package \citep{lightkurve} we extracted photometry from small apertures surrounding both V928 Tau and CFHT-BD-Tau 7, confirming the dimming event in fact originates from V928 Tau.
The asymmetric eclipse, after subtracting a stellar variability model and correcting for dilution due to the source binarity (as described in Section \ref{sec:lc}), is shown in Figure~\ref{fig:eclipse}.

\subsubsection{Photometry from Ground-Based Surveys}
\begin{figure}
    \centering
    \includegraphics[width=1\textwidth]{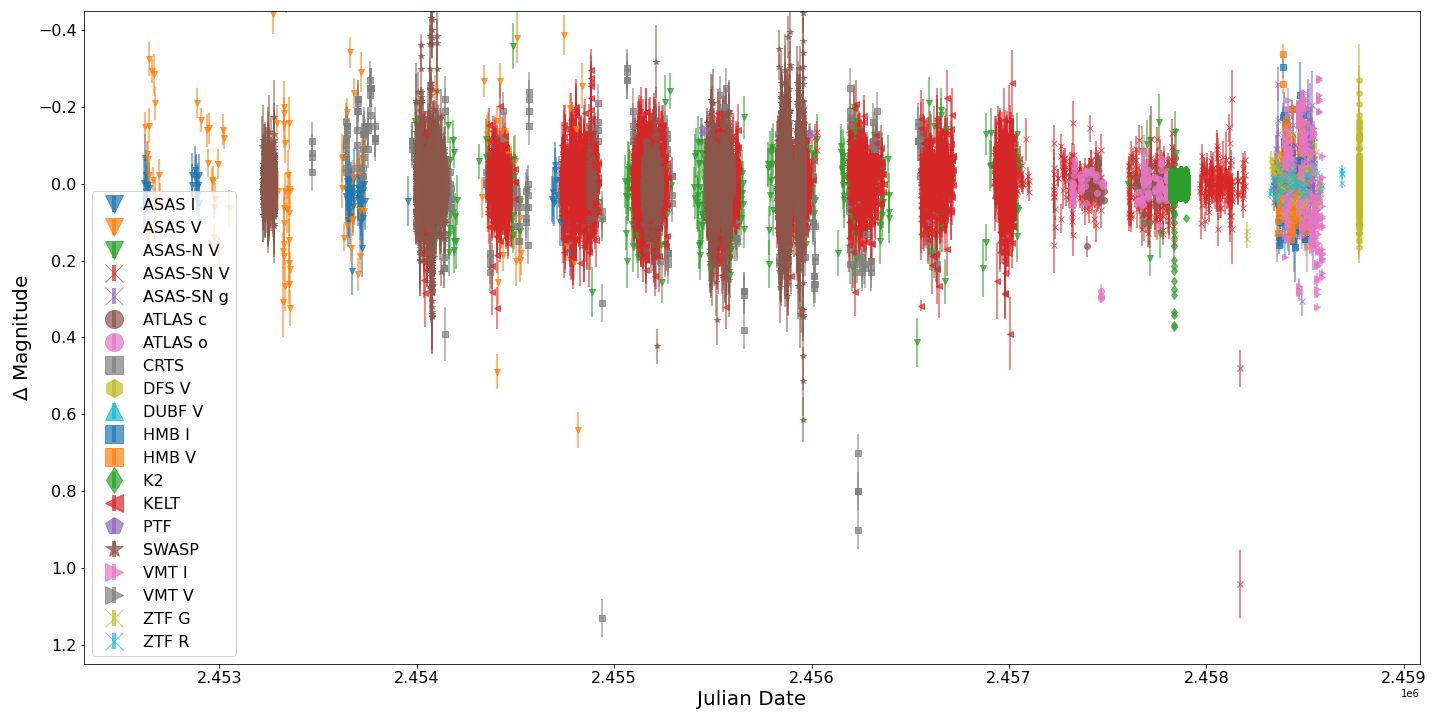}
    \caption{Time series photometry of V928 Tau from several time domain surveys. Note that data with magnitude errors exceeding 0.1 mag have been clipped for readability. Though there are several deep points not all of these are believable dips. The most believable dips are shown in Figure~\ref{fig:interesting_periods} after period folding the photometry.}
    \label{fig:all_phot}
\end{figure}

\begin{figure}
    \centering
    \includegraphics[width=\textwidth]{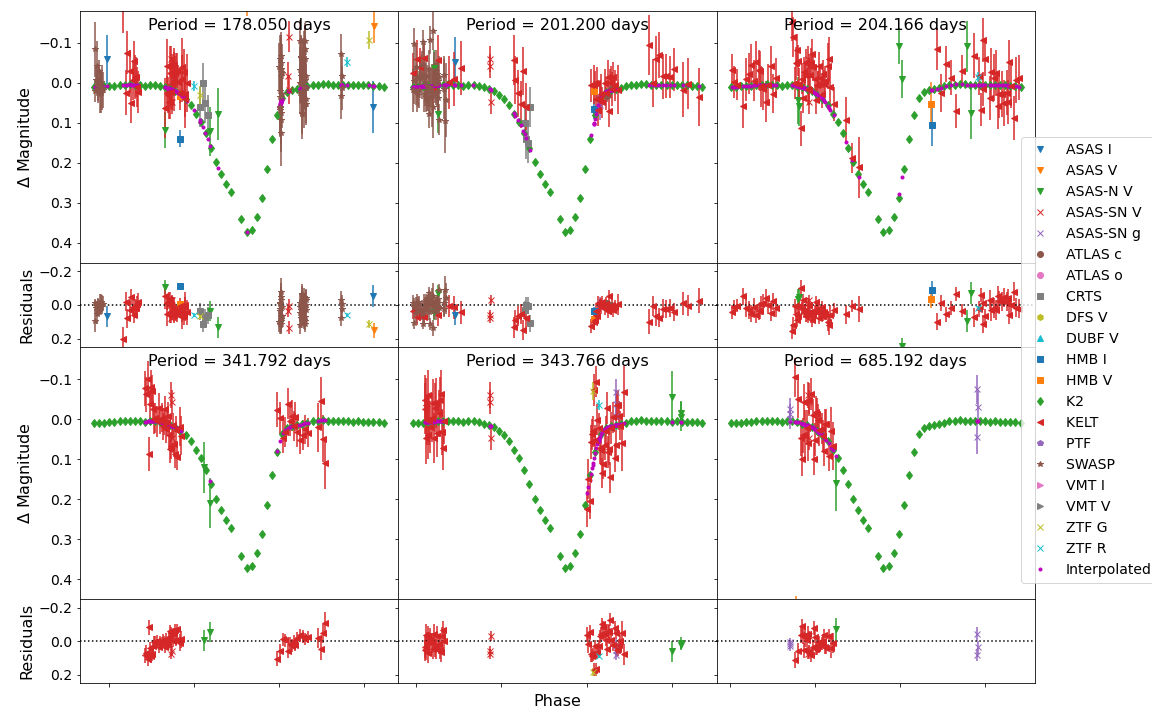}
    \caption{The photometry shown in Figure~\ref{fig:all_phot} is period folded with a high-resolution period grid and then visually inspected near the eclipse to determine the most interesting periods. Most interesting means that there is a suggestion of another eclipse taking place with the given period. Note that the phase labels have been removed from the plot as they provide no interesting information.}
    \label{fig:interesting_periods}
\end{figure}

To search for periodicity and long-term photometric variability of V928 Tau we supplemented the \ktwo\ data with photometry from various ground-based surveys (see Figure~\ref{fig:all_phot}).
Information on each survey is listed in Table~\ref{table:survey}, and though Figure~\ref{fig:all_phot} shows several brightness minima for V928 Tau, not all are believed to be real.
The most believable periods (visually determined), after applying period folding and removing stellar variation, are depicted in Figure~\ref{fig:interesting_periods}.

\input{table_ground_surveys}

The photometry we gathered originates from the following time-domain surveys. 
The All Sky Automated Survey \citep[ASAS,][]{Pojmanski:1997}, which consists of three separate telescopes at two locations, with a limiting magnitude of 13 mag and precision of 0.05 mag in $I$ band. 
The All Sky Automated Survey for Super-Novae \citep[ASAS-SN,][]{Shappee:etal:2014,Kochanek:etal:2017}, consists of five stations of four telescopes each, with a limiting magnitude of 17 mag. 
The Asteroid Terrestrial-impact Last Alert System \citep[ATLAS,][]{Tonry:etal:2018,Heinze:etal:2018}, consists of two telescopes with a limiting magnitude of about 19 mag. 
The Catalina Real-Time Transient Survey \citep[CRTS,][]{Drake:etal:2009}, consists of three telescopes with a limiting magnitude of 22 mag, and take data without a filter.
The Kilodegree Extremely Little Telescope \citep[KELT,][]{Pepper:etal:2007,KELT2012,KELT2018}, which consists of two telescopes designed to observe $V$ magnitudes between 7 and 11 mag with 1\% precision, but capable of observing stars down to $V$ = 14 mag.
The Palomar Transient Factory \citep[PTF,][]{Law:etal:2009,Rau:etal:2009}, which consists of one telescope for transient detection and one for photometric follow-up with a limiting magnitude of 20.6 mag in Mould-$R$ band.
The Super Wide-Angle Search for Planets \citep[SWASP,][]{Pollacco:etal:2006}, which consists of sixteen telescopes at two locations, designed to observe $V$ magnitudes between 7.0 and 11.5 mag with 1\% precision, but capable of observing stars down to $V$ = 15 mag.
The Zwicky Transient Facility \citep[ZTF,][]{Bellm:2014}, which expands on the PTF concept, consisting of a single telescope that has a limiting magnitude of 20.8 mag for ZTF $G$ band and 20.6 for ZTF $R$ band.
Data was also collected from amateur astronomers Franz-Josef Hambsch (HMB), Sjoerd Dufoer (DFS), Tonny Vanmunster (VMT) and the Astrolab Iris team (DUBF, Siegfried Vanaverbeke, Franky Dubois, Steve Rau and Ludwig Logie).

Data from the amateur astronomers was obtained through the American Association for Variable Star Observers (AAVSO) website\footnote{https://www.aavso.org/main-data}; ASAS-SN, ATLAS, CRTS, and ZTF surveys are publicly available from the project websites; the KELT light curve for V928 Tau was published in \citet{Rodriguez:etal:2017}; and the data from SWASP and ASAS are made publicly available for the first time here. 

We checked for additional photometric data from the DASCH digitized photographic plate archive (J. Grindlay, private communication), the HATNet Exoplanet Survey (J. Hartman, private communication), the Next Generation Transit Survey (E. Gillen, private communication) and Evryscope (N. Law, private communication).
Unfortunately, data for V928 Tau from these projects and surveys either does not presently exist or has not been processed.

A high-cadence light curve of $<$0.5 day duration from the Optical Monitor on board the XMM-Newton satellite was published in \citet{Audard:etal:2007}. 
Not surprisingly, no eclipses were detected over that brief period. 

\subsection{Spectroscopy: Keck-I/HIRES}
\label{sec:spectroscopy}
We observed V928 Tau with the HIRES spectrograph \citep{Vogt:etal:1994} at the Keck-I telescope on 2017-10-05 UT, 2017-12-10 UT and 2018-11-03 UT.
For the first and third epochs, our HIRES reduction and analysis procedures are identical to those discussed in \citet{David:etal:2019}.
The radial velocity of the spatially and spectrally unresolved pair was determined from cross-correlation \citep{Tonry:Davis:1979} of the spectrum with those of standard stars \citep{Nidever:etal:2002} observed on the same night. 
The two measurements are formally consistent with one another.  
However, a better constraint on radial velocity variations comes from cross correlating the observations with one another; this reveals an upper limit of $<$ 1~\kms on the difference in radial velocity at the two epochs.
The cross correlations are somewhat flat-topped, but it was not possible to separate the signals from what is likely the two stellar components at approximately the same velocity.
From the first epoch spectrum we also determined the sky-projected rotational velocity by artificially broadening a spectral standard using the \citet{Gray:2005} broadening profile, as well as the equivalent widths of the H$\alpha$, H$\beta$, and $\mathrm{Ca}_{\mathrm{\,II}}$ H $\&$ K lines, all of which are observed in emission. %
The third epoch spectrum used a redder setting of HIRES and enabled us to measure \ion{Li}{1}, and also to note that the \ion{Ca}{2} ``infrared'' triplet lines have sub-continuum core emission. 

Our second epoch of HIRES observations were reduced and analyzed following the California Planet Search procedures outlined in \citet{Howard:etal:2010}. 
The radial velocity at this epoch was determined using the telluric A and B  absorption bands as a wavelength reference \citep{Chubak:etal:2012}.
While this method typically yields uncertainties of 0.1--0.3 \kms\ for slowly rotating stars, we determined an uncertainty of 3.5 \kms\ from the RMS of 3/4 of the spectral segments used to calculate the RV.

\subsection{Gaia DR2}
Despite the brightness of V928 Tau, neither a parallax nor proper motions are available for the source from \gaia\ DR2 (ID 147799312239072000). 
This is likely a consequence of the source's binarity, as indicated by the large values of the goodness of fit statistic of the astrometric model with respect to along-scan observations (137.8864) and the excess astrometric noise (4.218 mas, 3690$\sigma$). 
\gaia\ DR2 did, however, publish a radial velocity estimate with large relative uncertainty: \vrad\, = 7.71 $\pm$ 6.50~\kms.
V928 Tau's low mass companion CFHT-Tau-7 ([MDM2001] CFHT-BD-Tau 7 = 2MASS J04321786+2422149), associated as mentioned in \citet{Kraus:Hillenbrand:2009b}, at 18\arcsec\, separation is \gaia\ DR2 147799209159857280.
\gaia\ DR2 reports parallax $\varpi$ = 8.0534\, $\pm$\, 0.1915 mas and proper motion \pmra, \pmdec\,  = 6.255, -22.196 $\pm$ (0.302, 0.233) \masyr.

\subsection{High-Resolution Imaging: Keck-II/NIRC2}
\label{sec:hires}
\begin{figure}
    \centering
    \includegraphics[width=\textwidth]{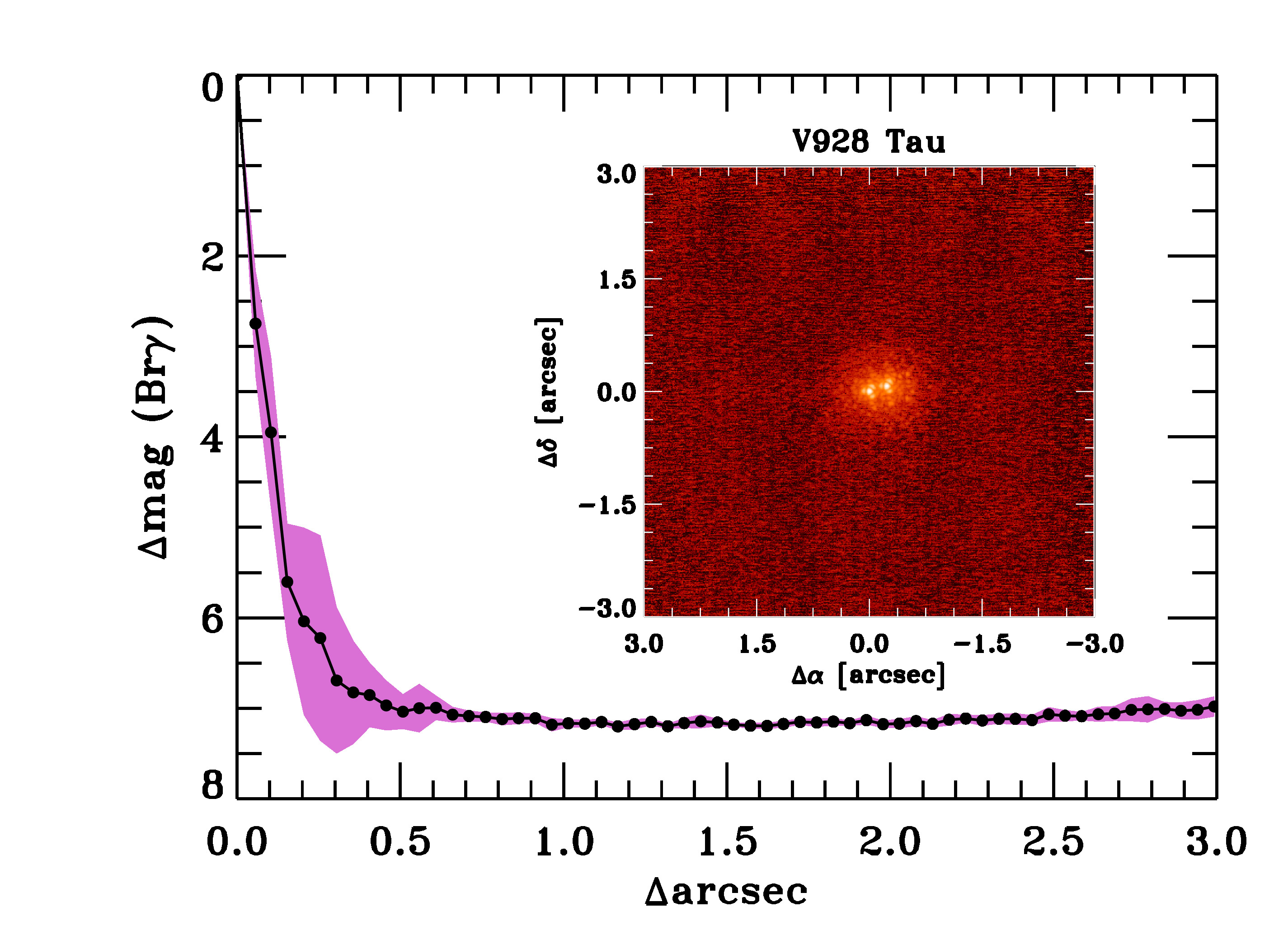}
\caption{Companion sensitivity for the Keck adaptive optics imaging in $\mathrm{Br}\gamma$.  The black points represent the 5$\sigma$ limits and are separated in steps of 1 image FWHM ($\sim 0.05$\arcsec); the purple shading represents the azimuthal RMS dispersion of the sensitivity. The wide dispersion in the 3--7 $\lambda / D$ range is due to the directional dependence of the sensitivity due to the presence of the close secondary (V928 Tau B). The inset image is of the target clearly showing the resolved binary.}\label{fig:ao_contrast}   
\end{figure}

We observed V928 Tau with infrared high-resolution adaptive optics (AO) imaging at Keck Observatory.  
The Keck Observatory observations were made with the NIRC2 instrument on Keck-II behind the natural guide star AO system. 
The observations were made on 2017-09-11 UT in the standard 3-point dither pattern that is used with NIRC2 to avoid the left lower quadrant of the  detector which is typically noisier than the other three quadrants. 
The dither pattern step size was $3\arcsec$ and was repeated twice, with each dither offset from the previous dither by $0.5\arcsec$.  
The observations were made in the narrow-band $\mathrm{Br}\gamma$ filter $(\lambda_o = 2.1686\,\mu \mathrm{m};\ \Delta\lambda = 0.0326\,\mu\mathrm{m}$) with an integration time of 2 seconds with one coadd per frame for a total of 18 seconds on target and in $J_{cont}$ $(\lambda_o = 1.2132\, \mu\mathrm{m};\ \Delta\lambda = 0.0198\,\mu\mathrm{m}$) with an integration time of 5 seconds with one coadd per frame for a total of 45 seconds on target. 
The camera was in the narrow-angle mode with a full field of view of $\sim10\arcsec$ and a pixel scale of approximately $0.0099442\arcsec$ per pixel.
The final combined dithers have a resolution of 0.049\arcsec\ in $\mathrm{Br}\gamma$ and 0.043\arcsec\ in $J_{cont}$. 
The Keck AO observations clearly show the binary in both filters, with the stars having a difference in magnitude of $\Delta K = 0.069\pm0.006$ mag and $\Delta J = 0.122 \pm 0.014$ mag. 
The observation also allows us to add another astrometric point to the emerging orbit for the stellar binary.
There are no additional stellar companions brighter than about $\Delta K (\mathrm{Br}\gamma) \approx 7$ magnitudes (5$\sigma$) compared to the primary to within a resolution of 0.1\arcsec\ ($\sim 14$ au, see Figure \ref{fig:ao_contrast}).

The sensitivities of the final combined AO image were determined by injecting simulated sources azimuthally around the primary target every $45^\circ$ at separations of integer multiples of the central source's FWHM \citep{furlan2017}. 
The brightness of each injected source was scaled until standard aperture photometry detected it with $5\sigma $ significance. 
The resulting brightness of the injected sources relative to the target set the contrast limits at that injection location. 
The final $5\sigma $ limit at each separation was determined from the average of all of the determined limits at that separation.
The uncertainty on the 5$\sigma$ limit was set by the RMS dispersion of the azimuthal slices at a given radial distance. 
The sensitivity curve is shown in Figure \ref{fig:ao_contrast} along with an inset image zoomed to the primary target showing no other companion stars.

\section{Light Curve Analysis}
\label{sec:lc}
As observed by \ktwo, V928 Tau is an unresolved, nearly equal brightness binary.  As such, the true eclipse depths are deeper by a factor dependent on the optical flux ratio and on which component is being eclipsed. 
Given the fact that the two stars are of similar spectral type, mass and radius, we take the limiting case where both stars are identical. 
\subsection{Rotational Modulation}
\begin{figure}
    \centering
    \includegraphics[width=\textwidth]{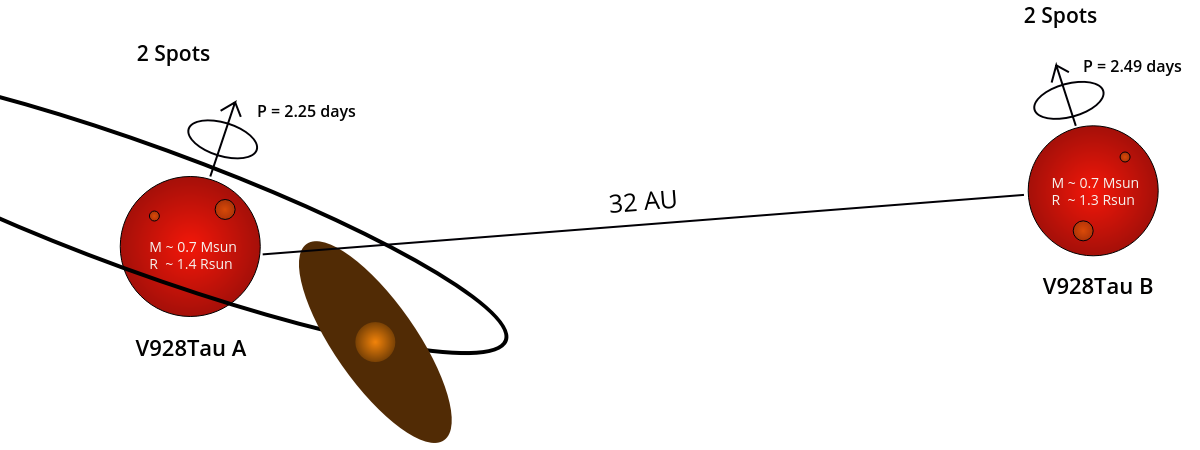}
    \caption{Interpretation of the \ktwo\ light curve of V928 Tau. The beating pattern arises from rotational modulation of each star (due to star spots) with very similar periods. The deep, asymmetric eclipse is likely caused by a companion with a disk that is elliptical due to projection effects. Note: figure is not to scale or a sky projection (line of sight not necessarily into the page), rotational periods might correspond to opposite stars and the proposed companion may orbit the other star.}
    \label{fig:cartoon}
\end{figure}

\begin{figure}
    \centering
    \includegraphics[width=\textwidth]{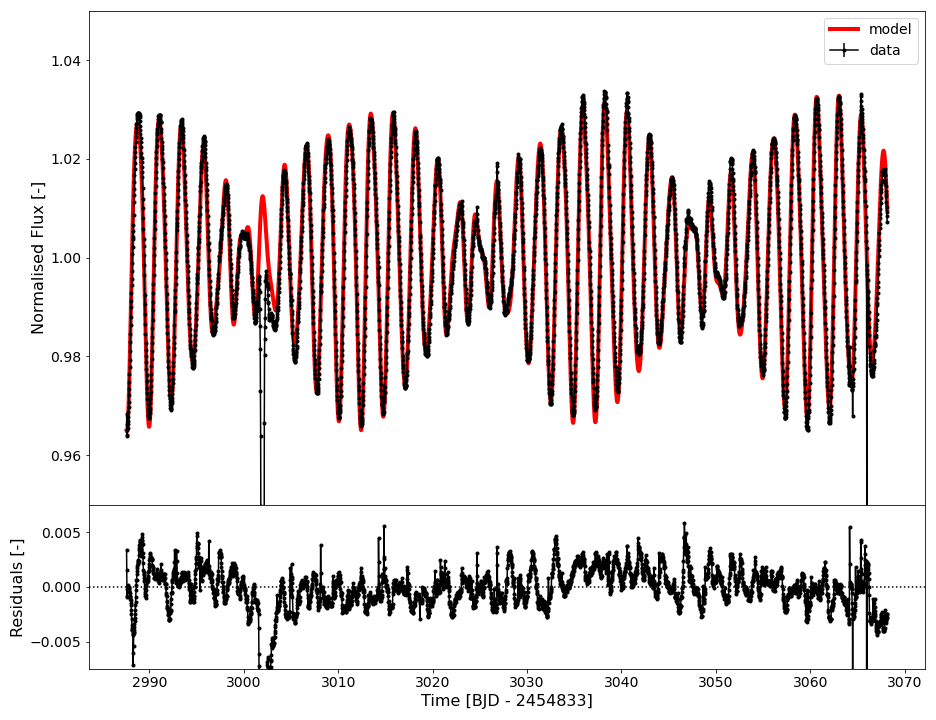}
    \caption{\textit{Top}: light curve for V928 Tau obtained from the \texttt{everest 2.0} pipeline with the MCMC model for the stellar modulations superimposed. \textit{Bottom}: residuals of the fit. }
    \label{fig:stellar_var_model}
\end{figure}

We interpret the  brightness modulations as originating from starspots on the surfaces of the binary components, and the beating pattern as arising from the nearly equal rotation periods of the two stars (see Figure~\ref{fig:cartoon}). 
Using a linear least-squares fit we remove the linear trend ($m = 0.00001422,\, c = 0.99973174$).
Using the Lomb-Scargle algorithm \citep{Lomb:1976, Scargle:1982} we find four significant sinusoidal periods at 1.130, 1.245, 2.249, and 2.485 days.
We note that if we accept a 1.0\% discrepancy (i.e. the percentage offset between a perfect harmonic, in other words, an integer ratio): 2.249 and 2.485 days are the two independent fundamental periods, with 1.130 and 1.245 being the respective first harmonics. 
To determine the amplitudes and phases of these modulations, we use the Levenberg-Marquardt Least-Squares algorithm \citep{10.2307/43633451, 10.2307/2098941} removing elements one by one.
We start with a linear trend with slope $m$ and $y$-intercept $c$, then two sinusoids, then again two sinusoids which have amplitudes $a_x$, periods $P_x$ and phases $\theta_x$, where $x = 1, 2, 3, 4$.
The least squares fit provides an initial guess for the MCMC simulation, and we run 250 chains with 10,000 links and a burn-in of 2,000 steps.
The results of the MCMC optimization are summarized in Table~\ref{table:stellar_var} and plotted in Figure~\ref{fig:stellar_var_model}. 
Note that there is no significant linear trend ($m = 0.00000224,\, c = 1.00022059$).
$P_1$ (2.250 days) and $P_2$ (2.482 days) contain the largest power and are interpreted to be the probable rotation periods of the two stars, which are very similar to the rotation periods found by \cite{Rebull:2020:arxiv}.
$P_3$ and $P_4$ are the first harmonic of $P_1$ and $P_2$ respectively (half periods), which are phase shifted w.r.t. the fundamental periods producing the asymmetric features in the beating light curve.
The exact physical reasons for this, whether it is a specific distribution of starspots, differential rotation or a combination of the two, is not relevant to this study as it is focused on characterising the eclipse, and the stellar variation model residuals are small ($< 0.5\%$)
Examining the ground-based data does not convincingly confirm or reject the stellar variation model determined by the \ktwo\ data.
This is likely due to the low amplitude of the modulation, the relatively high uncertainties on ground measurements and the likelihood that the stellar activity (spots and phages) evolves with time on the surface of the star.
\input{table_stellar_variations.tex}

\subsection{Eclipse Fitting}
\label{subsec:harddisk}
We note that the eclipse observed in the \ktwo\ photometry is most likely the result of the occulter eclipsing a single component of V928 Tau.
Usually, one would have to make separate models for the transit along either star, but given the fact that the two stars are of similar spectral type, mass and radius, we take the limiting case where both stars are identical.
In this case we simply double the fluctuations about the median (of one) and obtain the de-blended light curve of V928 Tau A/B (see Figure~\ref{fig:eclipse}).
After correction for the true eclipse depth we find that the eclipse depth exceeds 50\%. 

Other systems that show similar lop-sided eclipses include EE Cephei, the similar $\epsilon$ Aurigae and TYC 2505-672-1.
These are the only known long-period eclipsing binary star systems with obscurations caused by a large dust disk surrounding one of the components.
EE Cephei has not been observed directly, but extensive modeling was done by \citet{Galan:2012} and later tested with an international observing campaign by \citet{Pienkowski:2020:ArXiv}.
$\epsilon$ Aurigae on the other hand was observed directly using Georgia State University's Center for High Angular Resolution Astronomy Interferometer \citep[CHARA,][]{CHARA} using the Mid Infra-Red Combiner \citep[MIRC,][]{MIRC} and modelled extensively \citep{Kloppenborg:etal:2010,Kloppenborg:etal:2015}.
\citet{Rodriguez:etal:2016} found TYC 2505-672-1, an M-type red giant that undergoes a $\sim$ 3.45 year long, near-total eclipse every 69.1 years due to a moderately hot ($\sim 8000$ K) object with a large circumstellar disk, by sifting through 120 years worth of light curves.
Other interesting systems are OGLE LMC-ECL-11893 \citep{Scott:etal:2014}, OGLE-BLG182.1.162852 \citep{Rattenbury:etal:2014}, which are modelled as circumstellar disks of an unseen companion transiting the primary.

The depth and asymmetry of the de-blended V928 Tau eclipse make it very unlikely that the eclipse is caused by another star, in an equatorial orbit.
Instead, this gives rise to the theory that the eclipse is caused by an inclined and tilted disk around an unseen object, which due to projection produces an elliptical occulter.

This disk is modelled as an azimuthally symmetric dust disk with radius, $R_d$, disk inclination, $i$, tilt (angle w.r.t. the orbital path), $\phi$, impact parameter (w.r.t. orbital path), $b$, and an opacity, $\tau$.
For matters of simplicity we assume the projection of the occulter can be modelled as a disk (no gap between body and disk, or companion bulge).
To model the eclipse, the linear limb-darkening parameter, $u$, of the star and the transverse velocity of the disk, $v_t$, are required (\rstar\ is needed to convert $v_t$ from $R_* \, \mathrm{day}^{-1}$ to \kms). 
The models for $u$ are dependent on the effective temperature, \teff, metallicity, [Fe/H], surface gravity, $\log{g}$, and microturbulence velocity $v_{\mu t}$.
\cite{Tottle:Mohanty:2015} find that \teff\ = 3525 K.
\cite{Padgett:1996} and \cite{D'Orazi:etal:2011} find that [Fe/H] of stars in the Taurus Auriga association are near solar ($< 0.1$), so we assume [Fe/H] = 0.0.
In the models of limb-darkening, $v_{\mu t}$ is restricted to 2 \kms, leaving $\log{g}$ to be inferred.
We can estimate $\log{g}$ using equation~\ref{eq:logg}, where $M_*$ and $R_*$ are the mass and radius of the star, respectively and $\log{g_\odot} \sim 4.44$ cm s$^{-2}$ (using IAU nominal values), based on the radii and masses given in Table~\ref{table:star}.
\begin{equation}
    \label{eq:logg}
    \log{g} = \log{g_\odot} + \log{\frac{M_*}{M_\odot}} - 2 \log{\frac{R_*}{R_\odot}}
\end{equation}
Corrections due to rotational velocity of the stars are negligible as they are a small fraction of the break-up velocity ($\sim 13\%$).
We use the \texttt{jktld} fortran code developed by \cite{jktld} to linearly interpolate (\teff\ and $\log{g}$) the tables from \citet{Sing:2010} for values of $u$ for the \kepler\ bandpass in each of the four cases (V928 Tau A and B, with Dartmouth standard and magnetic models).
We take $u$ to be the average of these four cases giving $u = 0.7220$. 
Given $u$, we can ensure the $v_t$ predicted by the MCMC sampling algorithm is physical by following the method of \citet{vanWerkhoven:etal:2014} to derive a lower limit for the speed of the occulting object by measuring the steepest time derivative of the light curve, $\dot{L}$, (the egress) and assuming the radius, $R$, for each star with $u$.
\begin{equation}
    \label{eq:vanwerkhoven}
    v_t = \dot{L} R \pi \left(\frac{2 u - 6}{12 - 12 u + 3 \pi u}\right)
\end{equation}
Using these values of $u$, the sizes of each star and the luminosity slope of the egress, $\dot{L} = 4.19 \ L_* \, \text{day}^{-1}$, we obtain a lower limit of $v_{t,A} = 65.5$ \kms, $v_{t,B} = 61.7$ \kms, which is consistent with the best-fit $v_t$.
This corresponds to $\sim 5.91\,R_*$.

As there are likely many acceptable configurations, we try to find the smallest disk that could cause the eclipse. 
The reason being that this can provide lower mass limits on the companion, and can constrain the disk size, in the most intuitive way. 
We do this in two ways: modelling a partially transmitting disk, which is preferentially opaque ($\tau$ from $0.5 - 1.0$) and a fully opaque disk ($\tau = 1$).

To perform the modelling of the elliptical occulter we use a modified version of the {\texttt{pyPplusS}} code developed by \cite{10.1093/mnras/stz2556}.
This code produces light curves in physical space, i.e. it determines the eclipse depth based on the physical area that has been blocked by the occulter (which can be a planet, disk, or planet disk/ring system combination of which we use the disk model).
This produces photometric points based on the geometry and location of the occulter w.r.t. the host star as well as the limb-darkening model of the star, which in this case we simplify to the linear model with parameter $u$.
Note further that this code works in units of stellar radii, which permits us to ignore the choice of star and the uncertainties on the radii.
However, to produce a light curve it is necessary to convert the spatial domain to the temporal domain by introducing $v_t$ and fitting for the time of maximum occultation, $\delta t$, with respect to $BJD = 2457835$. 

We start off by initialising a set of 1,000 chains for 3,500 links with the initial bounds as described in Table~\ref{table:mcmc_bounds} and bind the probability by the parameter bounds.
We further check to make sure that all the initial chains produce a transit (otherwise it might be too far removed to converge to a given solution), and as a final check we check if the system is physical.
The maths and limits to determine whether or not a set of model parameters produces a physical disk is described in detail in Section~\ref{subsec:periodicity}, but the basic concept is as follows.
A disk is considered physical if for a given companion mass, $M_p$, (we use 80 \mjup, which is an upper limit for the deuterium burning limit), and a maximum apastron distance (we use 3.2 au as this is 10\% of the binary separation, which fulfills a stability criterion) corresponds to $R_d < 0.3 \, R_H$, where $R_H$ is the Hill radius of the companion.
With a fixed $M_p$, and the choice of a star ($M_*$ and $R_*$, we use V928 Tau B), this becomes solely dependent on $v_t$.
\input{table_mcmc_bounds}

\begin{figure}
    \centering
    \includegraphics[width=\textwidth]{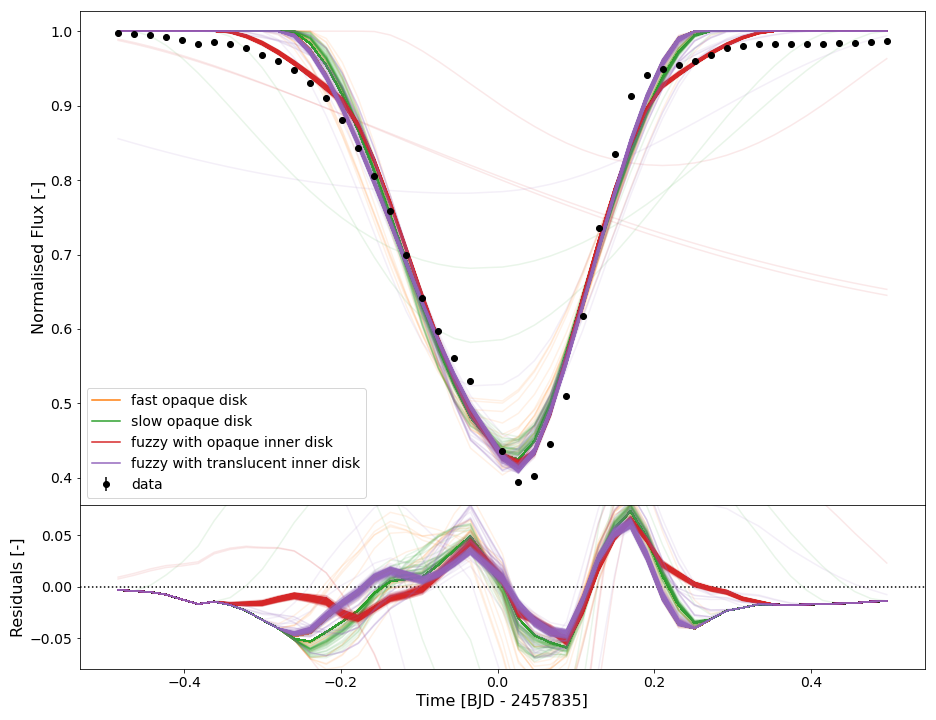}
    \caption{\textit{Top}: Results of the MCMC sampling algorithm for the different local minima for the single component opaque disk model in orange ($v_t > 8 \,R_*\,\mathrm{day}^{-1}$) and green ($v_t < 8 \,R_*\,\mathrm{day}^{-1} \ \& \ R_d < 1.5 \,R_*$) and the two component fuzzy disk model in red ($\tau > 0.9$) and purple ($\tau < 0.9$) for the eclipse of V928 Tau with the best-fit parameters summarized in Table~\ref{table:eclipse_models}. \textit{Bottom}: Residuals of the MCMC samples. Note that the "fast opaque disk" and the "fuzzy with translucent inner disk" practically overlap.} 
    \label{fig:eclipse_models}
\end{figure}

Performing the MCMC optimisation reveals several local minima for the eclipse solutions, namely a high velocity set ($v_t > 8\, R_* \, \mathrm{day}^{-1}$, 381 chains, burn-in 500 links) and a low velocity set with small disk radii ($v_t < 8\, R_* \, \mathrm{day}^{-1}$ and $R_d < 1.5 \,R_*$, 472 chains, burn-in 1,000 links).
We also find that in both cases the opaque disk produces a better fit than the translucent disk, so we scrap the translucent solutions.
The results of the MCMC optimisation are summarised in Table~\ref{table:eclipse_models} (Opaque Fast and Opaque Slow columns) and visualised in Figure~\ref{fig:eclipse_models}.
Note that the errors displayed in the table are on the MCMC distribution itself, whereas the systematic errors are much larger.
Examples in these errors include, uncertainties in $u$, $R_*$, the assumption that the two stars are identical so the de-blended light curve is as depicted in Figure~\ref{fig:eclipse}.
Also consider the fact that this model does not include scattering of light and other such processes that would influence the shape of the light curve.
\subsection{Two Component Disk Model}
\label{subsec:fuzzydiskmodel}

We also attempt a two component fuzzy disk model where we add two parameters to the model, namely the thickness of the second (edge) component, $t_e$, and its opacity $\tau_e$.
Note that the total radius of the fuzzy disk is the sum of $R_d$ and $t_e$.
We run the same procedure described in section~\ref{subsec:harddisk}, with these two additional parameters.
Performing the MCMC optimisation reveals two local minima for the eclipse solutions, namely a fuzzy with opaque inner disk ($\tau > 0.9$, 457 chains, burn-in 350 links) solution (red) and a fuzzy with translucent inner disk ($\tau < 0.9$, 543 chains, burn-in 350 links) solution (purple).
The results of the MCMC sampling are summarised in Table~\ref{table:eclipse_models} (Fuzzy Opaque and Fuzzy Translucent columns) and visualised in Figure~\ref{fig:eclipse_models}.
Due to the significantly higher $v_t$, we adopt the single, low velocity, small radius opaque disk model.

\input{table_eclipse_fits}

\subsection{Periodicity}
\label{subsec:periodicity}
The eclipse observed by \ktwo\ does not repeat over the baseline of those observations. 
Since the eclipse occurs during the first half of the \ktwo\ campaign, a lower limit on a potential period is obtained from $P_{orb} > t_{K2,\mathrm{end}} - t_\mathrm{ecl} - t_\mathrm{dur}$, where $t_{K2,\mathrm{end}}$ is the final timestamp from \ktwo, $t_\mathrm{ecl}$ is the eclipse midpoint, and $t_\mathrm{dur}$ is the eclipse duration. 
In this case, the period of the candidate eclipsing companion must be $P_{orb} > 66$ days. 

We construct models on the orbit and periodicity of the proposed companion.
We initially assume that $v_t$ corresponds to a circular orbit, leading to a semi-major axis $a \sim$ 0.1 au and $P_{orb} < 66$ days - given that no other eclipse is seen, this rules out circular orbits for the occulter.
The orbit must therefore be eccentric and to investigate possible orbits we assume that $v_t = v_{peri}$ and explore a grid containing the mass of the companion, $M_p$, and $P_{orb}$.

We determine an upper bound for $M_p$ given that the spectra of the both components of the binary are nearly identical and that there is no obvious tertiary companion in the high spatial resolution images (see Figure~\ref{fig:ao_contrast}).
To do this we take the upper mass limit of substellar objects, i.e. the deuterium burning limit.
Despite the fact that more recent studies by \citet{Baraffe:etal:2015} and \citet{Forbes:Loeb:2019} show that the deteurium burning limit is 73-74 \mjup, we take the older upper limit of 80 \mjup\ determined by \citet{Saumon:Marley:2008} to be inclusive of higher masses.
\citet{Quarles_2020} find that for a companion to remain bound to its host in a binary star system with $a_{bin}$, the orbit of the companion must have $a/a_{bin} < 0.08$ for a prograde orbit. 
For a retrograde orbit this fraction increases to $0.10$.
By taking the upper limit of 10\%, which results in $a = 3.2$ au, we find that $P_{orb} = 2.8$ years for a circular orbit. 
Thus, $P_{orb}$ is run from 66 days to 2.8 years.

\begin{figure}
    \centering
    \includegraphics[width=\textwidth]{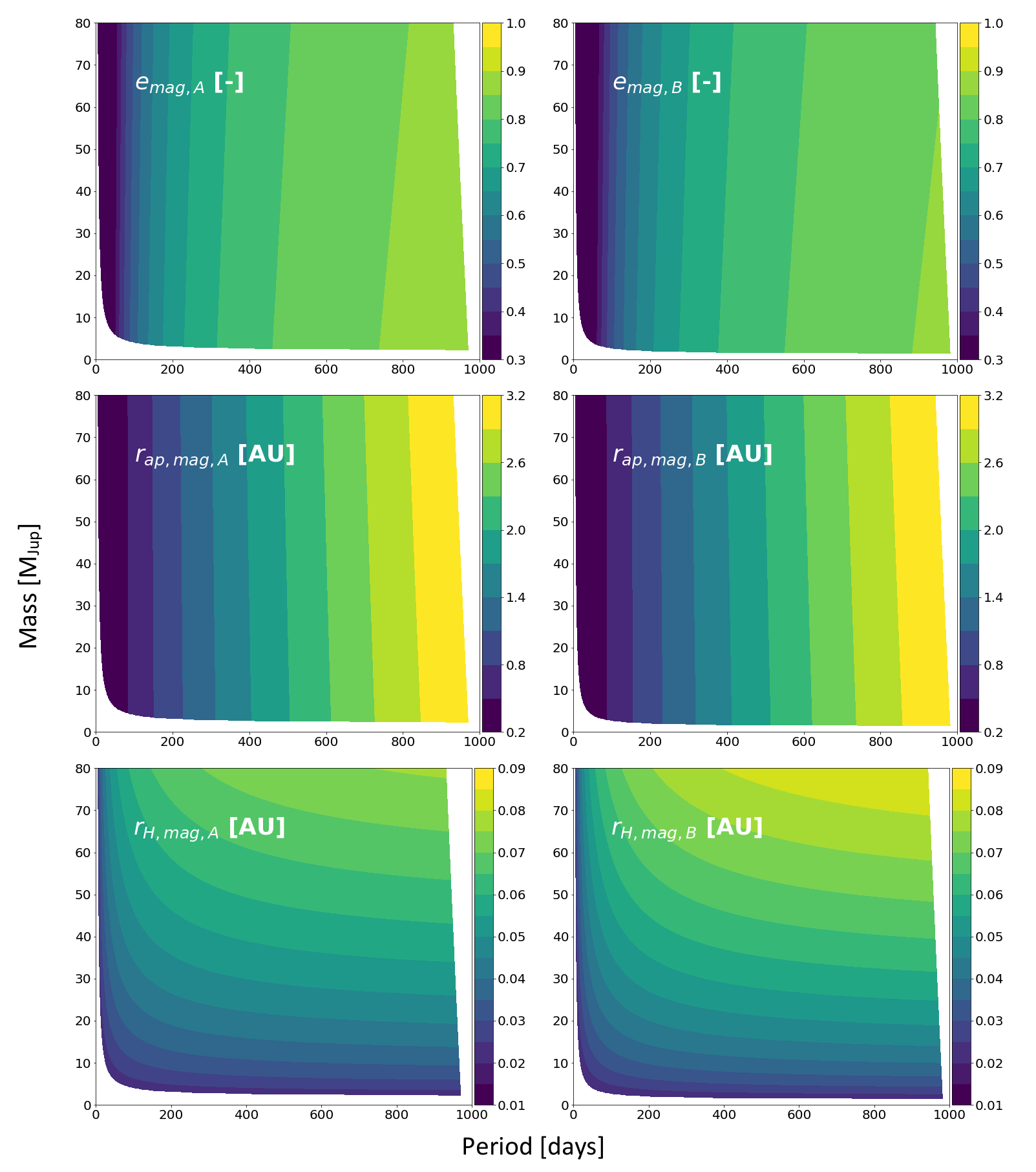}
    \caption{Parameter spaces mapped out for the hard disk model with a companion mass, $M_p$, and orbital period, $P_{orb}$, grid for the magnetic models of V928 Tau A and B (see Table~\ref{table:star}). \textit{Left} and \textit{right} panels show properties if the companion orbits either V928 Tau A or B, respectively. From \textit{top} to \textit{bottom}, the properties mapped are eccentricity $e$ [-], apastron distance $r_{ap}$ [au] and the Hill radius $r_H$ [au]. The bottom of each parameter map is masked out due to the $r_H$ constraint ($r_{disk} < 0.3\, r_H$). The right side of each parameter map is masked out due to the $r_{ap}$ constraint ($r_{ap} < 0.10 \, a_{bin}$).}
    \label{fig:props_mag}
\end{figure}

With a fixed mass of the host, $M_*$, and $v_t = v_{peri}$; a grid of $M_p$ and $P_{orb}$ we can determine the eccentricity, $e$, the periastron distance, $r_{peri}$, which we require to determine the Hill radius, $r_H$, and the apastron distance, $r_{ap}$.
We do this as follows. We use Kepler's third law to determine the semi-major axis, $a$.
Using $a$ we can determine $e$ by isolating it from the equation for $v_{peri}$ (equation~\ref{eq:vp}, where $\mu = G (M_* + M_p)$).
\begin{equation}
    \label{eq:vp}
    v_{peri} = \sqrt{\frac{(1 - e) \mu}{(1 + e) a}}
\end{equation}
Using $e$ and $a$ we can determine $r_{peri}$ and $r_{ap}$, and finally we use $r_{peri}$ to estimate $r_H$ as shown in equation~\ref{eq:rh}.
\begin{equation}
    \label{eq:rh}
    r_H = r_{peri} \sqrt[3]{\frac{M_p}{3 (M_* + M_p)}}
\end{equation}
For a disk to be stable over extended periods of time its radius, $r_{disk} < 0.3 \, r_H$.
Given that the companion will spend a significant fraction of its orbital period at $r_{ap}$, we constrain the system by requiring that $r_{ap} < 3.2$ au as the orbit must remain stable.
This method with the given constraints reveals that the opaque fast model requires $M_p > 50 M_\mathrm{jup}$, the fuzzy opaque and fuzzy translucent models require $M_p > 75 M_\mathrm{jup}$.
We thus adopt the opaque slow model for which the parameter maps are shown in Figure~\ref{fig:props_mag} for the magnetic models of V928 Tau.

This figure shows that the $r_{ap}$ constraint limits $P_{orb}$ to $\sim 1,000\, \mathrm{days}$ for the magnetic models.
The $r_H$ constraint carves out the region at the bottom of the maps so the minimum $M_p$ increases as $P_{orb}$ decreases.

\section{Astrometric Analysis}

We used astrometry compiled from the literature and our newly acquired data point from NIRC2 to fit plausible Keplerian orbits to the data. 
The relative astrometry between V928 Tau A \& B are found in Table~\ref{table:astrometry}.

\begin{figure}
        \centering
        \includegraphics[width=\textwidth]{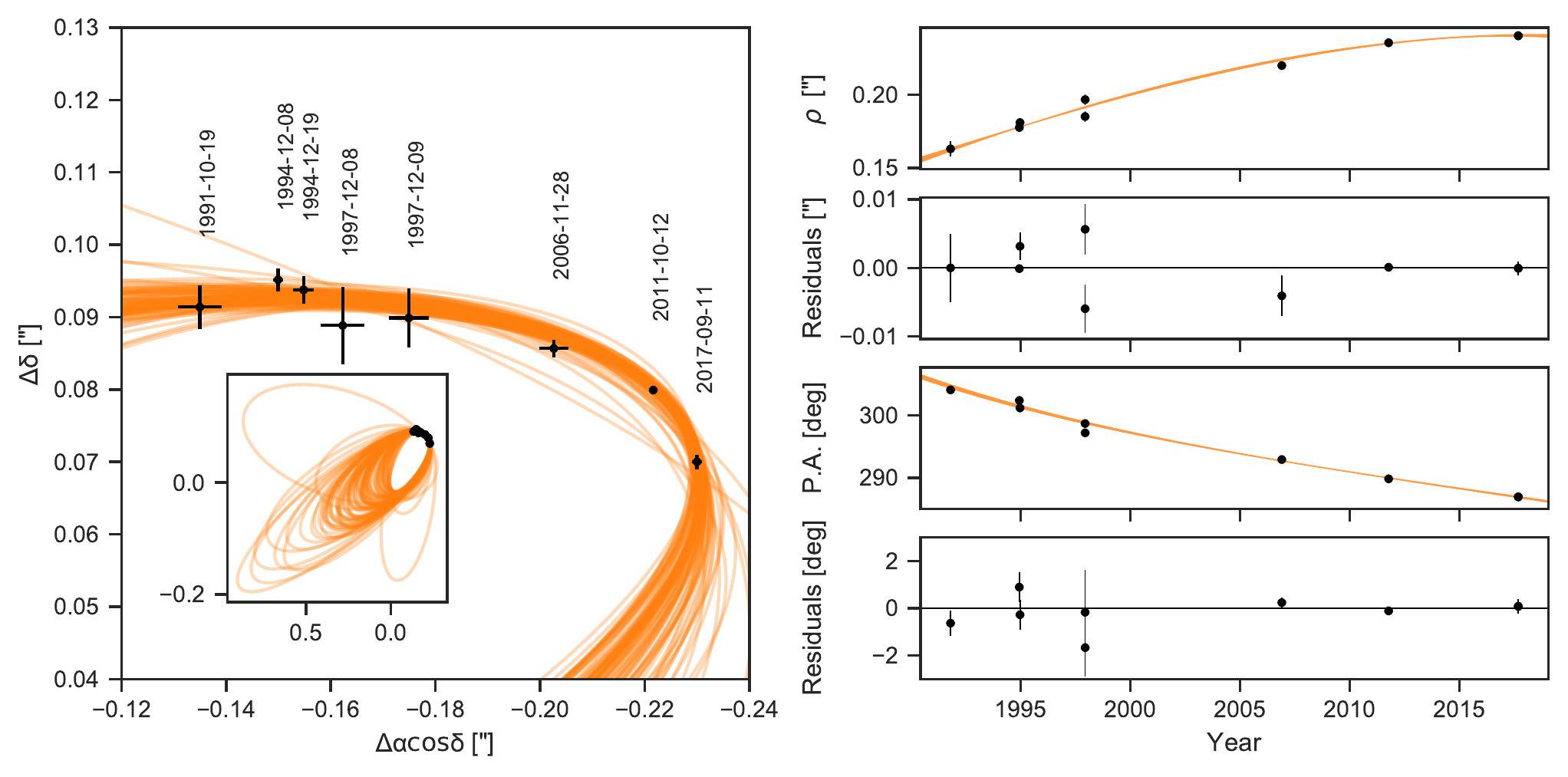}
        \caption{Relative astrometry of V928 Tau A+B. \textit{Left:} The orbit of V928 Tau B as it appears on the sky, with V928 Tau A at the origin (outside the plot range). Random draws from the posterior are shown by the orange curves. The inset panel shows the full orbit. \textit{Upper right:} Time series of angular separation measurements (above) and residuals (below). \textit{Lower right:} Time series of position angle measurements (above) and residuals (below).}
        \label{fig:astrometry}
    \end{figure}

Our analysis closely follows the \texttt{exoplanet} \citep{exoplanet:exoplanet} tutorial available online.\footnote{\url{https://docs.exoplanet.codes/en/stable/tutorials/astrometric/}} 
After an initial optimization with 
\texttt{scipy.optimize.minimize} using the BFGS method to find the maximum a posteriori solution we sampled from the posterior distribution using \texttt{exoplanet} and \texttt{PyMC3} \citep{salvatier2016}. 
The free parameters of the model were the log of the orbital period ($\log P$), $p = (\Omega + \omega)/2$, $m = (\Omega - \omega)/2$, eccentricity ($e$), cosine of the inclination ($\cos{i}$), a phase angle, the projected semi-major axis in arcseconds ($a$), the parallax ($\varpi$), and jitter terms for the angular separation and position angle ($\log s_{\rho}$ and $\log s_{\theta}$). 
We assumed Gaussian priors on the total system mass ($\mu$=1.4~\msun, $\sigma$=0.1~\msun) and the parallax ($\mu$=8.0534~mas, $\sigma$=0.1915~mas). 
Since the data only cover a small fraction of the orbit, we sampled only to get a coarse understanding of the posterior distribution and did not sample until convergence (for example, the Gelman-Rubin statistic for the orbital period was 1.02).
We used 4 chains with 10,000 links and a burn-in of 5,000 steps, for a final chain length of 20,000.
Nevertheless, from this preliminary sampling we determined that 68\% (99.7\%) highest posterior density interval for the orbital period is 73--171 (67--597) years. 
Our lower bound on the orbital period of the binary ($>$67 years at 99.7\% confidence) is somewhat larger than the minimum period of 58 years found by \citet{Schaefer:etal:2014} using the same data without our most recent measurement. 
The time series astrometry and model fits drawn from the posterior are shown in Figure~\ref{fig:astrometry}. 
Future modeling with longer astrometric and radial velocity time series should better constrain the binary's orbit.

\input{table_astrometry.tex}

\input{table_astrometric_fit.tex}

\section{Discussion}
\label{sec:discussion}
We find a model (opaque slow) for a companion orbiting either V928 Tau A or B with a surrounding dust disk with size $\sim 0.99 \,R_*$, which is $\sim 1.36$ \rsun\ for V928 Tau A and $\sim 1.28$ \rsun\ for V928 Tau B, which is significantly smaller than the proposed dust disk for J1407 \citep{Kenworthy_2015} and EPIC 204376071 \citep{Rappaport:etal:2019}, nevertheless significantly larger than the expected radius for Roche rings (e.g. Saturn's rings). 
This is the case for both absolute size and relative size (compared to $R_*$).
We kept the model as simple as possible, but a larger number of degrees of freedom (i.e. a ring system, with varying opacities, or an attenuating disk) can always result in a better fit.
Another feature to note in the model is the difficulty in modelling the ``wings'' of the eclipse.
These can be partially justified with a transition from transparent to opaque along the disk edge, but this results in an unphysically large edge.
One could imagine that an attenuating disk model could solve this with a size between the single component hard disk model and the two component fuzzy disk model and could thus be a physical disk.

We argue that the companion should be on a highly eccentric orbit and relatively high mass, up to a brown dwarf (80 \mjup).
The implied non-zero eccentricity seems to support a trend as we find that J1407 ($e > 0.7$) and EPIC 204376071 ($e > 0.33$) both require eccentric orbits to explain the lack of other eclipses in their light curves, implying that the companion disk plays an important role in planetary dynamics and could play a major role in the dynamical evolution of planet formation.
If this trend is discovered in other systems it implies that the circumplanetary disk may play a role in the migration of the companion.
\cite{2020AJ....159...63B} show that directly imaged brown dwarfs have a preference for higher eccentricities (in line with J1407) and gas giants have a strong preference for small eccentricities (EPIC 204376071 is in conflict with this result).
\citet{Winn_2015} show that distribution of $e$ tends to focus on small values for short periods, broadening at longer periods.

Further discoveries of other disk and ring systems, and confirmation of the orbital periods of these known systems will resolve this observation.
We selected a disk with the highest opacity for the occulter, but of course there are a family of companion disks, which have the same eclipse profile but are larger in diameter.
These run into the issue of stability within the Hill sphere, as explored by \cite{Rieder_2016} for J1407.
Follow up observations of this particular system would allow us to further characterise it, determine its composition through multi-filter observations (and thus the grain size distribution), along with high-resolution spectra to determine the chemical composition of the surrounding disk and companion.
V928 Tau is a particularly hard system to detect as the eclipse length is approximately half a day, allowing the eclipse to be hidden by the diurnal cycle. Further modelling of the stellar variation spanning the whole baseline of observations (including the full activity cycle of the star) could reveal hints of another eclipse providing potential periods for predictions of the next transit event.
Follow up observations that either confirm the existence of the occulters, or that detect other eclipses in these systems, will help us understand the nature of these intriguing systems.

\section{Acknowledgements} 

We thank Dan Foreman-Mackey, Ian Czekala, Sarah Blunt for helpful discussions on the astrometric modeling.
This paper includes data collected by the Kepler mission and obtained from the MAST data archive at the Space Telescope Science Institute (STScI). 
Funding for the Kepler mission is provided by the NASA Science Mission Directorate.
STScI is operated by the Association of Universities for Research in Astronomy, Inc., under NASA contract NAS 5–26555.
Part of this research was carried out at the Jet Propulsion Laboratory, California Institute of Technology, under a contract with the National Aeronautics and Space Administration (80NM0018D0004).
TJD and EEM gratefully acknowledge support from the Jet Propulsion Laboratory Exoplanetary Science Initiative
and NASA award 17-K2GO6-0030.
Some of the data presented herein were obtained at the W. M. Keck Observatory from telescope time allocated to the National Aeronautics and Space Administration through the agency's scientific partnership with the California Institute of Technology and the University of California. 
The Observatory was made possible by the generous financial support of the W. M. Keck Foundation. 
The authors wish to recognize and acknowledge the very significant cultural role and reverence that the summit of Maunakea has always had within the indigenous Hawaiian community. 
We are most fortunate to have the opportunity to conduct observations from this mountain.
This work has made use of data from the Asteroid Terrestrial-impact Last Alert System (ATLAS) project. 
ATLAS is primarily funded to search for near earth asteroids through NASA grants NN12AR55G, 80NSSC18K0284, and 80NSSC18K1575; byproducts of the NEO search include images and catalogs from the survey area. 
The ATLAS science products have been made possible through the contributions of the University of Hawaii Institute for Astronomy, the Queen's University Belfast, the Space Telescope Science Institute, and the South African Astronomical Observatory.

\software{EVEREST (v2.0; \citet{Luger:etal:2016,Luger:etal:2018}), lightkurve (\citet{lightkurve}), exoplanet (\citet{exoplanet:exoplanet}), PyMC3 (\citet{salvatier2016}), Scipy (\citet{2020SciPy-NMeth}, Matploltib \citep{Hunter:2007}, Astropy \citep{astropy:2013,astropy:2018}, Numpy \citep{numpy}, jktld \citep{jktld}}

\bibliography{main}

\end{document}

%% file: table_star.tex
\startlongtable
\begin{deluxetable*}{lcl}
    \tablecaption{Parameters of V928 Tau \label{table:star}}
    \tablecolumns{3}
    \tablewidth{-0pt}
    \tabletypesize{\scriptsize}
    \tablehead{\colhead{Parameter} &
            \colhead{Value} & 
            \colhead{Reference}\\
            \colhead{} &
            \colhead{(primary, secondary)} &
            \colhead{}}
    \startdata
        \textit{Kinematics and position}\\
        R.A., J2000 (hh mm ss) & 04 32 18.88 & \gaia\ DR2\\
        Dec., J2000 (dd mm ss) & +24 22 26.71 & \gaia\ DR2\\
        $\mu_\alpha$ (\masyr) & 18.6 $\pm$ 5.1 & \citet{Zacharias:etal:2015}\\
        $\mu_\delta$ (\masyr) & -21.2 $\pm$ 5.1 & \citet{Zacharias:etal:2015}\\
        \vrad\, (\kms) & 15.38\,$\pm$\,0.16 & \citet{Nguyen:etal:2012}\\
        $\varpi$ (mas) & 8.0534 $\pm$ 0.1915 & \gaia\ DR2 -- CFHT-Tau-7\\ 
        Distance (pc) & 124 $\pm$ 3 & \gaia\ DR2 -- CFHT-Tau-7 \\ 
        \\
        \textit{Photometry}\\
        $u$ (mag)   & 18.000 $\pm$ 0.012 & SDSS DR12 \\
        $g$ (mag)   & 15.367 $\pm$ 0.004 & SDSS DR12 \\
        $r$ (mag)   & 14.772 $\pm$ 0.011 & SDSS DR12 \\
        $i$ (mag)   & 15.841 $\pm$ 0.014 & SDSS DR12 \\
        $z$ (mag)   & 12.619 $\pm$ 0.011 & SDSS DR12 \\
        $G$ (mag)   & 12.8122 $\pm$ 0.0018 & \gaia\ DR2 \\
        $G_\mathrm{BP}$ (mag)   & 14.3086 $\pm$ 0.0078 & \gaia\ DR2 \\
        $G_\mathrm{RP}$ (mag)   & 11.6026 $\pm$	0.0045 & \gaia\ DR2 \\
        $J$ (mag)   &  9.538 $\pm$ 0.020 & 2MASS \\
        $H$ (mag)   &  8.432 $\pm$ 0.021 & 2MASS \\
        $K_s$ (mag) &  8.106 $\pm$ 0.021 & 2MASS \\
        $W1$ (mag) & 7.906 $\pm$ 0.023 & \textit{WISE -- All-Sky} \\ 
        $W2$ (mag) & 7.804 $\pm$ 0.019 & \textit{WISE -- All-Sky} \\ 
        $W3$ (mag) & 7.717 $\pm$ 0.022 & \textit{WISE -- All-Sky} \\ 
        $W4$ (mag) & 7.705 $\pm$ 0.294 & \textit{WISE -- All-Sky} \\ 
        \\
        \textit{Deblended Photometry}\\
        $J$ (mag) & 10.23 $\pm$ 0.03, 10.35 $\pm$ 0.03 & this work\\ 
        $K_s$ (mag) & 8.82 $\pm$ 0.02, 8.89 $\pm$ 0.02 & this work\\ 
        \\
        \textit{Dereddened Photometry}\\
        $J_\mathrm{0}$ (mag) & 9.77 $\pm$ 0.05, 9.90 $\pm$ 0.05 & this work\\ 
        $K_\mathrm{s,0}$ (mag) & 8.64 $\pm$ 0.03, 8.67 $\pm$ 0.03 & this work\\ 
        \\        
        \textit{Physical properties}\\
        Spectral type & M0.8 $\pm$ 0.5 & \citet{Herczeg:Hillenbrand:2014}\\
        $A_V$ (mag) &  1.95 $\pm$ 0.2 & \citet{Herczeg:Hillenbrand:2014} \\
        $E(B-V)$ (mag) & 0.63 $\pm$ 0.07 & this work \\
        $T_\mathrm{spec}$ (K) & 3660 $\pm$ 70, 3660 $\pm$ 70 & this work \\
        $T_\mathrm{phot}$ (K) & 3610 $\pm$ 110, 3640 $\pm$ 110 & this work \\
        $\log(L_*/L_\odot)$ (dex) & -0.518 $\pm$ 0.031, -0.570 $\pm$ 0.032 & this work\\
        $R_*$ (\rsun) &  1.376 $\pm$ 0.059, 1.296 $\pm$ 0.056 & this work\\
        $M_*$ (\msun) & 0.70 $\pm$ 0.07, 0.70 $\pm$ 0.07 & this work -- DMM \\
                      & 0.45 $\pm$ 0.05, 0.46 $\pm$ 0.05 & this work -- DSM \\
        $\tau_*$ (Myr) & 5.8 $\pm$ 1.5,  6.9 $\pm$ 1.8 & this work -- DMM \\
                       & 2.5 $\pm$ 0.6,  3.0 $\pm$ 0.7 & this work -- DSM \\
        $v\sin{i_*}$ (\kms) & 29 $\pm$ 3  & this work - 2017 spectrum \\
        & 33.1 $\pm$ 1.2 & this work - 2018 spectrum \\
        & 34.2 $\pm$ 0.4 & \citet{Kounkel:etal:2019} \\
        & 31.6 $\pm$ 0.7 & \citet{Nguyen:etal:2012} \\
        & 18.8 $\pm$ 3.3 & \citet{Hartmann:Stauffer:1989} \\
        & 24.9 & \citet{Hartmann:etal:1986} \\
        EW(H$\alpha$) (\AA) & -0.95 & this work \\
        EW(H$\beta$) (\AA) & -0.89 & this work \\
        EW(\ion{Ca}{2} H) (\AA) & -8.9 & this work \\
        EW(\ion{Ca}{2} K) (\AA) & -13.4 & this work \\
        EW(\ion{Li}{1} 6707.8) (m\AA) & 658 & this work\\
         & 639 & \citet{Martin:etal:1994} \\
        $P_\mathrm{rot,1}$ (d) & 2.25 & this work \\
        $P_\mathrm{rot,2}$ (d) & 2.48 & this work \\
    \enddata
    \tablenotetext{}{
        References: 2MASS = \citet{2006AJ....131.1163S}; \gaia\ DR2 = \citet{GaiaDR2}; SDSS DR12 = \citet{Alam_2015}; \textit{WISE -- All-Sky} = \citet{2010AJ....140.1868W}. 
        DMM = Dartmouth Magnetic Models \citep{Feiden:2016}, DSM = Dartmouth Standard Models \citep{Dotter:etal:2008}.}
\end{deluxetable*}

%% file: table_kinematics.tex
\begin{deluxetable*}{llllll}[bh]
    \tablecaption{Astrometry for V928 Tau AB and Neighboring Tau IV Subgroup Members \label{table:kin}}
    \tablecolumns{6}
    \tablewidth{-0pt}
    \tabletypesize{\scriptsize}
    \tablehead{\colhead{ID} &
            \colhead{Catalog} & 
            \colhead{$\varpi$} & 
            \colhead{\pmra} & 
            \colhead{\pmdec}\\
            \colhead{} &
            \colhead{} & 
            \colhead{(mas)} &
            \colhead{(\masyr)} &
            \colhead{(\masyr)}}
    \startdata
        V928 Tau                & HSOY     &  ...  & 5.816\,$\pm$\,2.130 & -29.200\,$\pm$\,2.096\\
        V928 Tau                & GPS1     &  ...  & 6.398\,$\pm$\,1.823 & -16.593\,$\pm$\,1.532\\
        V928 Tau                & PPMXL    &  ...  & 5.8  \,$\pm$\,4.5   & -29.8  \,$\pm$\,4.5  \\
        \hline  
        2MASS J04321786+2422149 & \gaia\ DR2 &  8.0534\,$\pm$\,0.1915 & 6.255\,$\pm$\,0.302 & -22.196\,$\pm$\,0.233\\
        FY Tau                  & \gaia\ DR2 &  7.6798\,$\pm$\,0.0710 & 6.651\,$\pm$\,0.135 & -21.855\,$\pm$\,0.116\\	
        FZ Tau                  & \gaia\ DR2 &  7.6908\,$\pm$\,0.0746 & 7.121\,$\pm$\,0.143 & -21.497\,$\pm$\,0.106\\	
        Haro 6-13               & \gaia\ DR2 &  7.6653\,$\pm$\,0.1879 & 5.017\,$\pm$\,0.317 & -21.378\,$\pm$\,0.243\\
        HK Tau A                & \gaia\ DR2 &  7.5005\,$\pm$\,0.0924 & 4.464\,$\pm$\,0.152 & -22.961\,$\pm$\,0.116\\
        HK Tau B                & \gaia\ DR2 &  5.1023\,$\pm$\,1.5260 & 0.369\,$\pm$\,2.520 & -27.032\,$\pm$\,2.032\\
        2MASS J04325026+2422115 & \gaia\ DR2 & 11.8560\,$\pm$\,2.4075 & 7.042\,$\pm$\,4.285 & -25.073\,$\pm$\,3.452\\	
        MHO 8                   & \gaia\ DR2 &  7.7979\,$\pm$\,0.2219 & 6.369\,$\pm$\,0.390 & -20.474\,$\pm$\,0.289\\
        \hline
        Tau IV                  & L09       &  7.14\,$\pm$\,0.51   & 5.5 \,$\pm$\,1     & -21.9 \,$\pm$\,1   \\
        median (Tau IV-V928)    & this work &  7.69\,$\pm$\,0.06   & 6.13\,$\pm$\,0.36  & -22.03\,$\pm$\,0.70\\
        mean (Tau IV-V928)      & this work &  7.36\,$\pm$\,0.36   & 6.31\,$\pm$\,0.77  & -22.20\,$\pm$\,0.52\\
    \enddata
    \tablenotetext{}{
        Mean is Chauvenet clipped mean. Uncertainties in mean are standard error. Uncertainties in median are uncertainty in true median. References: 
        \gaia\ DR2  = \citet{GaiaDR2}. 
        GPS1 = \citet{GPS12017}. 
        HSOY = \citet{HSOY2017}. 
        L09 = \citet{Luhman:etal:2009}. 
        PPMXL = \citet{PPMXL2010}. }
\end{deluxetable*}

%% file: table_rvs.tex
\begin{deluxetable*}{lll}
    \tablecaption{Radial Velocities of V928 Tau A+B \label{table:rvs}}
    \tablecolumns{3}
    \tablewidth{-0pt}
    \tabletypesize{\scriptsize}
    \tablehead{\colhead{Date} &
            \colhead{RV} & 
            \colhead{Reference} \\
            \colhead{(JD)} &
            \colhead{(\kms)} &
            \colhead{}}
    \startdata
        ... & 18.3\,$\pm$\,2.0\tablenotemark{a} & \citet{Hartmann:etal:1986} \\
        ... & 15.38\,$\pm$\,0.16 & \citet{Nguyen:etal:2012} \\
        ... & 7.71\,$\pm$\,6.50 & \citet{GaiaDR2}\\	
        ... & 16.1\,$\pm$\,0.23 & \citet{Kounkel:etal:2019} \\
        ... & 18                & \citet{Zhong:etal:2019} \\
        2458032.11194  & 16.0\,$\pm$\,1.8\tablenotemark{b} & this work \\
        2458097.888854 & 14.4\,$\pm$\,3.5\tablenotemark{b} & this work \\
        2458425.83663 & 17.9\,$\pm$\,2.8\tablenotemark{b} & this work \\
    \enddata
    \tablenotetext{a}{The RV uncertainty for the \citet{Hartmann:etal:1986} measurement has been estimated from Table 1 of that work.}
    \tablenotetext{b}{Radial velocities derived from spatially unresolved spectroscopy of the blended binary.}
\end{deluxetable*}

%% file: table_ground_surveys.tex
\begin{deluxetable*}{lccccccc}[ht]
    \tablecaption{Ground Survey Information \label{table:survey}}
    \tablecolumns{8}
    \tablewidth{-0pt}
    \tabletypesize{\scriptsize}
    \tablehead{\colhead{Survey} &
            \colhead{Filter} & 
            \colhead{$n_{tel}$} &
            \colhead{Baseline} & 
            \colhead{$n_{phot}$} &
            \colhead{Pixel-Scale} &
            \colhead{Field of View} &
            \colhead{Reduction}\\
            \colhead{} &
            \colhead{} &
            \colhead{} &
            \colhead{(days)} &
            \colhead{} &
            \colhead{(\arcsec\ pix$^{-1}$)} &
            \colhead{(deg$^2$ \, cam$^{-1}$)} &
            \colhead{(reference)}}
    \startdata
        ASAS$^a$  & $I$   & 1 -- 3 & 2213 & 121 & 14.2 & 6.0, 77.4     & \citet{Pojmanski:1997}\\
                  & $V$   &        & 2213 & 133   &      &               &   \\
                  & $V^b$ &        & 3859 & 508   &      &               &   \\
        ASAS-SN   & $V$   & 8      & 2505 & 664   & 8.0  & 20.3          & \citet{Kochanek:etal:2017}\\
                  & $g$   & 12     &  196 & 201   &      &               &   \\
        ATLAS     & $c$   &  8     &  527 & 132   & 1.9  & 28.9          & \citet{Heinze:etal:2018}\\
                  & $o$   &        & 510  & 143   &      &               &   \\
        CRTS      & $-$   & 3      & 3168 & 412   & 2.5  & 8.0, 1.0, 4.2 & \citet{Drake:etal:2009}\\
        K2        & $K_p$ & 1      & 81   & 3900  & 4.0  & 110           & \citet{Luger:etal:2016,Luger:etal:2018}\\
        KELT      & $R^c$   & 2      & 2987 & 9888  & 23   & 676           & \citet{Siverd:etal:2012}\\
        PTF       & $R$   & 1      & 547  & 4     & 1.0  & 8.1           & \citet{Masci_2016}\\
        SWASP     & $V$   & 16     & 2740 & 33704 & 13.7 & 64            & \citet{Pollacco:etal:2006}\\
        ZTF       & $G$   & 1      & 370  & 63    & 1.0  & 47            & \citet{Masci_2018} \\
                  & $R$   &        & 363  & 67    &      &               &   \\
        \hline
        DFS       & $V$   & 1      & 1    & 194   & 2.0  & 0.6           & \citet{LesvePhotometry} \\
        DUBF      & $V$   & 1      & 46   & 63    & 1.9  & 0.1           & \citet{Meng:etal:2017} \\
        HMB       & $I$   & 2      & 143  & 113   & 2.1, 2.2  &  0.6, 0.7  & \citet{LesvePhotometry}\\
                  & $V$   &        & 143  & 118   &      &               &   \\
        VMT       & $I$   &        & 191  & 654   & 1.8  & 0.5           & \citet{LesvePhotometry}\\
                  & $V$   &        & 2    & 7     &      &           &   \\
    \enddata
    \tablenotetext{a}{Upgraded in 2002 from one to three telescopes in Chile.}
    \tablenotetext{b}{Telescope in Hawaii.}
    \tablenotetext{c}{non-standard, see \citet{Pepper:etal:2007}}
\end{deluxetable*}

%% file: table_stellar_variations.tex
\begin{deluxetable*}{cccccc}
    \tablecaption{\\ Sinusoidal Stellar Variations \label{table:stellar_var}}
    \tablecolumns{6}
    \tablewidth{-0pt}
    \tabletypesize{\scriptsize}
    \tablehead{\colhead{Mode} &
            \colhead{Amplitude} & 
            \colhead{Period} &
            \colhead{Phase} & 
            \colhead{Harmonic Modes} & 
            \colhead{Harmonic Discrepancy}\\
            \colhead{} &
            \colhead{(\%)} &
            \colhead{(days)} &
            \colhead{(rad)} &
            \colhead{} &
            \colhead{(\%)}}
    \startdata \centering
        1 & 2.0 & 2.250 & 1.379 & ...   & ...   \\
        2 & 1.1 & 2.482 & 1.671 & ...   & ...   \\
        3 & 0.1 & 1.130 & 1.352 & 1     & 0.91  \\
        4 & 0.3 & 1.245 & 1.456 & 2     & 0.63  \\
    \enddata
    \tablenotetext{}{Harmonic Modes indicate which modes are multiples of each other. Harmonic Discrepancy is the percentage off a perfect harmonic (e.g. with periods 1 and 2.1 days the harmonic discrepancy would be 10\%).}
\end{deluxetable*}

%% file: table_mcmc_bounds.tex
\begin{deluxetable*}{cccc}
    \tablecaption{\\ MCMC Boundaries \label{table:mcmc_bounds}}
    \tablecolumns{6}
    \tablewidth{-0pt}
    \tabletypesize{\scriptsize}
    \tablehead{\colhead{Parameter} &
            \colhead{Parameter Bounds} & 
            \colhead{Initial Walker Bounds} &
            \colhead{Units}}
    \startdata \centering
        $R_d$       & 0 -- 10    & 0 -- 5      & $R_*$  \\
        $b$         & -10 -- 10  & -5 -- 5     & $R_*$  \\
        $i$         & 0 -- 90    & 45 -- 90    & deg    \\
        $\phi$      & 0 -- 90    & 0 -- 90     & deg    \\
        $v_t$       & 5.9 -- 20    & 5.9 -- 10     & $R_*\,\mathrm{day}^{-1}$ \\
        $\delta t$  & -10 -- 10  & -0.5 -- 0.5 & day    \\
        $\tau$      & 0 -- 1     & 0.5 -- 1    &        
    \enddata
        \tablenotetext{}{Notes on the Parameter Bounds.\\ 
        1) Upper bound for $R_d$ has been deemed large enough.\\
        2) The bounds for $b$ are such that the disk must transit the star.\\
        3) Due to reflection symmetries caused by the combination of $b$ and $\phi$, $\phi$ is limited from 0$^\circ$ -- 90$^\circ$ instead of -180$^\circ$ -- 180$^\circ$. \\
        4) The lower bound for $v_t$ corresponds to the method discussed in \cite{vanWerkhoven:etal:2014}, with an upper bound deemed large enough.}
\end{deluxetable*}

%% file: table_eclipse_fits.tex
\begin{deluxetable*}{lcccc}
    \tablecaption{Eclipse Model Parameters \label{table:eclipse_models}}
    \tablecolumns{5}
    \tablewidth{-0pt}
    \tabletypesize{\scriptsize}
    \tablehead{
            \colhead{Parameter} & 
            \colhead{Opaque Fast} &
            \colhead{Opaque Slow} &
            \colhead{Fuzzy Opaque} &
            \colhead{Fuzzy Translucent}}
    \startdata
        $R_d$ [$R_*$] & 1.9392 $\pm$ 0.0005 & 0.9923 $\pm$ 0.0005 & 1.0017 $\pm$ 0.0166 & 2.2481 $\pm$ 0.0569  \\
        $t_e$ [$R_*$] &   $-$   &   $-$   & 1.7813 $\pm$ 0.0271 & 0.0420 $\pm$ 0.0599  \\
        $b$ [$R_*$] & 0.8519 $\pm$ 0.0007 & -0.2506 $\pm$ 0.0002 & -0.3171 $\pm$ 0.0110 & 0.8670 $\pm$ 0.0370  \\
        $i$ [$^\circ$] & 67.11 $\pm$ 0.02 & 56.78 $\pm$ 0.03 & 61.1 $\pm$ 0.8 & 65.0 $\pm$ 0.6  \\
        $\phi$ [$^\circ$] & 24.83 $\pm$ 0.02 & 41.22 $\pm$ 0.05 & 44 $\pm$ 2 & 40 $\pm$ 2  \\
        $v_t$ [$R_*\,\mathrm{day}^{-1}$] & 9.135 $\pm$ 0.002 & 6.637 $\pm$ 0.002 & 8.2 $\pm$ 0.1 & 9.1 $\pm$ 0.2  \\
        $v_{t,A}$ [$\mathrm{km s}^-1$] & 101.22 $\pm$ 0.02 & 73.53 $\pm$ 0.02 & 90 $\pm$ 2 & 101 $\pm$ 2  \\
        $v_{t,B}$ [$\mathrm{km s}^-1$] & 95.33 $\pm$ 0.02 & 69.26 $\pm$ 0.02 & 85 $\pm$ 2 & 95 $\pm$ 2  \\
        $\delta t$ [$\mathrm{day}$] & -0.0586 $\pm$ 0.0001 & 0.0099 $\pm$ 0.0000 & 0.0179 $\pm$ 0.0009 & -0.0629 $\pm$ 0.003  \\
        $\tau$ [-] & 1.0 & 1.0 & 0.997 $\pm$ 0.003 & 0.67 $\pm$ 0.02  \\
        $\tau_e$ [-] &   $-$   &   $-$   & 0.163 $\pm$ 0.005 & 0.17 $\pm$ 0.05  \\
    \enddata
    \tablenotetext{}{
        The total size of the disk is the sum of $R_d$ and $t_e$.}
    \tablenotetext{}{
        For the conversion of $v_t$ to km s$^{-1}$ we use $R_* = 1.376 \, R_{\odot}$ for $v_{t,A}$ and $R_*= 1.296 \, R_\odot$ for $v_{t,B}$ corresponding to the radii of V928 Tau A and B.}
\end{deluxetable*}

%% file: table_astrometry.tex
\begin{deluxetable*}{llllll}[ht]
    \tablecaption{Relative Astrometry of V928 Tau A \& B \label{table:astrometry}}
    \tablecolumns{6}
    \tablewidth{-0pt}
    \tabletypesize{\scriptsize}
    \tablehead{\colhead{Date} &
            \colhead{$\rho$} & 
            \colhead{P.A.} &
            \colhead{Flux ratio} & 
            \colhead{Band} &
            \colhead{Reference}\\
            \colhead{(UT)} &
            \colhead{(\arcsec)} &
            \colhead{(deg.)} &
            \colhead{} &
            \colhead{} &
            \colhead{}}
    \startdata
        \nodata & 0.18 $\pm$ 0.01 & 300 $\pm$ 4 & 0.88 $\pm$ 0.03 & $K$ & \citet{Leinert:etal:1993} \\ 
        1991-10-19 & 0.163 $\pm$ 0.005 & 304.1 $\pm$ 0.5 & \nodata & $K$ & \citet{Ghez:etal:1995} \\
        1994-12-08 & 0.1776 $\pm$ 0.0002 & 302.4 $\pm$ 0.6 & 1.0 $\pm$ 0.1 & $V$ & \citet{Simon:etal:1996} \\
        1994-12-19 & 0.181 $\pm$ 0.002 & 301.2 $\pm$ 0.6 & \nodata & $K$ & \citet{Ghez:etal:1995} \\
        1997-12-08 & 0.1851 $\pm$ 0.0035 & 298.7 $\pm$ 1.8 & 1.009 $\pm$ 0.002 & $L$ & \citet{White:Ghez:2001} \\
        1997-12-09 & 0.1967 $\pm$ 0.0037 & 297.2 $\pm$ 1.2 & 1.055 $\pm$ 0.037 & $K$ & \citet{White:Ghez:2001} \\
        2006-11-28 & 0.220 $\pm$ 0.003 & 292.92 $\pm$ 0.09 & 0.9728 $\pm$ 0.0089 & $K'$ &  \citet{Kraus:Hillenbrand:2012} \\ 
        2011-10-12 & 0.23562 $\pm$ 0.00012 & 289.827 $\pm$ 0.031 & 0.9751 $\pm$ 0.0058 & $K_\mathrm{cont}$ & \citet{Schaefer:etal:2014} \\
        2017-09-11 & 0.24042 $\pm$ 0.00099 & 286.93 $\pm$ 0.24  & 0.938 $\pm$ 0.026 & $K_s$ & this work \\           &                       &                   & 0.893 $\pm$ 0.025 & $J$ & this work \\
    \enddata
    \tablenotetext{}{
        We have added 180$^\circ$ to the position angles reported by \citet{Ghez:etal:1995}, \citet{Simon:etal:1996}, and \citet{White:Ghez:2001} for consistency with the other surveys. A precise date was not given for the \citet{Leinert:etal:1993} measurement so we did not include it in our astrometric analysis.}
\end{deluxetable*}

%% file: table_astrometric_fit.tex
\begin{deluxetable*}{lrrrr}[ht]
    \tablecaption{Results of astrometric fit.\label{table:astrometric_fit}}
    \tablecolumns{5}
    \tablewidth{-0pt}
    \tabletypesize{\scriptsize}
    \tablehead{\colhead{Variable} &
            \colhead{Mean} & 
            \colhead{Std. dev.} &
            \colhead{HPD 3\%} & 
            \colhead{HPD 97\%}}
    \startdata
    \textit{Sampled} \\
    \hline 
    $\log{P}$        &       10.897 &      0.510 &       10.175 &       11.823 \\
    $\log{s_\rho}$   &       -9.563 &      3.206 &      -15.652 &       -5.122 \\
    $\log{s_\theta}$  &       -5.952 &      1.329 &       -7.450 &       -4.175 \\
    $a$ ('')       &        0.266 &      0.103 &        0.149 &        0.465 \\
    $p$ (rad)          &        1.668 &      0.305 &        1.185 &        2.048 \\
    $m$ (rad)          &       -0.006 &      0.357 &       -0.514 &        0.635 \\
    phase (rad)      &        0.171 &      2.085 &       -2.899 &        3.126 \\
    $\cos{i}$    &       -0.267 &      0.116 &       -0.511 &       -0.148 \\
    $e$         &        0.554 &      0.169 &        0.336 &        0.928 \\
    $\varpi$ (mas)       &        8.028 &      0.192 &        7.672 &        8.391 \\    
    \hline
    \textit{Derived} \\
    \hline
    $P$ (yr)          &      171.090 &    107.930 &       70.860 &      370.898 \\
    $a$ (au)           &       33.119 &     12.904 &       18.773 &       57.695 \\
    $t_\mathrm{peri}$       &  2454177.322 &  16129.837 &  2428221.544 &  2471110.331 \\
    $\omega$ (deg)       &       95.913 &     33.232 &       36.497 &      139.630 \\
    $\Omega$ (deg)      &       95.283 &     18.507 &       76.146 &      105.711 \\
    $i$ (deg)        &      105.711 &      7.334 &       98.549 &      120.722 \\
    \enddata
\end{deluxetable*}